# CARMENES as an Instrument for Exoplanet Research

José A. Caballero
Centro de Astrobiología, CSIC-INTA, ESAC Campus, Camino Bajo del Castillo s/n, 28692 Villanueva de la Cañada, Madrid, Spain
e-mail: caballero@cab.inta-csic.es

Walter Seifert, Andreas Quirrenbach
Landessternwarte, Zentrum für Astronomie der Universität Heidelberg, Königstuhl 12, 69117 Heidelberg, Germany
e-mails: aquirren@lsw.uni-heidelberg.de, w.seifert@lsw.uni-heidelberg.de

Pedro J. Amado
Instituto de Astrofísica de Andalucía (IAA-CSIC), Glorieta de la Astronomía s/n, 18008 Granada, Spain
e-mail: pja@iaa.es

Ignasi Ribas
Institut de Ciències de l'Espai (ICE-CSIC), Campus UAB, c/ Can Magrans s/n, 08193 Bellaterra, Barcelona, Spain and Institut d'Estudis Espacials de Catalunya (IEEC), c/ Gran Capità 2–4, 08034 Barcelona, Spain
e-mail: iribas@ice.csic.es

Ansgar Reiners
Institut für Astrophysik und Geophysik, Georg-August-Universität, Friedrich-Hund-Platz 1, 37077 Göttingen, Germany
e-mail: ansgar.reiners@phys.uni-goettingen.de

**Abstract** CARMENES stands for Calar Alto high-Resolution search for M dwarfs with Exoearths with Near-infrared and optical Échelle Spectrographs. CARMENES took six years from a concept to the start of operations, and a couple more years of initial data collection until the first science publication, but now is revolutionising our knowledge on exoplanets and their stars in our immediate vicinity. Here we describe what CARMENES is: (*i*) an ultra-stabilised two-channel spectrograph at an almost dedicated 3.5 m telescope in southern Spain that covers in high spectral resolution and without big gaps from 0.52 μm to 1.71 μm; (*ii*) a science project aimed at comprehensively searching for and studying planetary systems with nearby, bright, M-dwarf hosts, but that also investigates transiting planets around other stars; and (*iii*) the German-Spanish consortium that designed and built the instrument and that has operated it under guaranteed and legacy time observations.

# Introduction

At the beginning of this millennium, Hatzes et al. (2003) had confirmed the long-period exoplanet around the bright star γ Cephei, which had been claimed fifteen years earlier by Campbell et al. (1988); the planetary system around the millisecond pulsar PSR B1257+12, discovered by Wolszczan & Frail (1992), had become even weirder; Charbonneau et al. (2000) had open the new window of transiting planets and, therefore, the characterisation of their atmospheres and bulk properties; and 51 Pegasi b, discovered by Mayor & Queloz (1995), was just the tip of an iceberg, which quickly became an ice field plenty of exoplanets of all radii and masses. This avalanche was thanks to, e.g., the technological and scientific competition between teams at both sides of the Atlantic, which pointed their spectrographs and photometers to the sky. Some used iodine gas cells for the wavelength calibration of their radial-velocity (RV) surveys, other used hollow cathode lamps. Some used tiny telescopes (or rather commercial cameras with telephoto lenses), other used the *Hubble Space Telescope*. But virtually all of them focused on solar-like stars, i.e., F- and G-type dwarfs in the solar neighbourhood.

The first light of the High Accuracy Radial velocity Planet Searcher (HARPS; Mayor et al. 2003) on the ESO 3.6 m Telescope was in February 2003. Soon after, HARPS was the first spectrograph to beat the 1 m/s boundary in RV precision. This "magic number", 1 m/s, became afterwards the scientific requirement of many spectrographs worldwide. These new projects wanted to go one step ahead by discovering less and less massive planets, down to the Earth-mass regime… And in the habitable zone! However, with the achievable RV precision and as a consequence of the variation of an exoplanet's RV semi-amplitude as a function of its host star mass ($K \propto M_*^{-2/3}$), one should focus on target stars with masses much lower than that of the Sun. Unfortunately for spectrograph users in early 2000s, which only covered in high resolution the blue and green parts of the optical, such low-mass stars were red M dwarfs. Unless they are extremely close, at merely a few parsecs, M dwarfs are too faint for optical, high-resolution spectrographs at 2-4 m-class telescopes because of their low luminosities and the bulk of their spectral energy distribution peaking in the near infrared. As a result, some teams started developing ideas of near-infrared high-resolution spectrographs able to perform RV surveys for Earth-mass planets in the habitable zone of M dwarfs.

Such RV surveys of M dwarfs faced some obstacles at the beginning of their long-distance race. At the time when the exoplanet business took off, people motivated their limitations to F and G stars with the argument that M dwarfs cannot host habitable planets because the formers are active and the later are tidally locked. As a result, frequent showers of steriliser radiation from the stellar coronae were expected to pour on the illuminated hemispheres of rocky planets, thought to be barren lands that lost their atmosphere and, therefore, any trace of liquid water. In the early 2000s, RV precision was also very low in M dwarfs because of the very low flux bluewards of 0.7 μm, and the spectral range redwards was essentially uncharted terrain. This changed slowly when some people looked for binary brown dwarfs, and when CCDs were used to study M dwarfs at wavelengths up to the silicon cutoff, where lots of photons and some isolated spectral lines live, although at very low efficiencies. While others, including some of the authors, had looked for habitable planets around M (and L) dwarfs with a number of techniques, it was not until Scalo et al. (2007) and Tarter et al. (2007) published their seminal reviews on the topic when M dwarfs moved back into focus of RV surveys.

In the second half of 2000s there were a bunch of RV spectrographs planned to operate in the near infrared with precision enough to beat the "magic number" of 1 m/s, inaccessible to single-order spectrographs such as CRIRES at the Very Large Telescope. One of the new long-distance runners was NAHUAL (Martín et al. 2004), which was planned to cover from 0.9 μm to 2.4 μm with a spectral resolution $R \sim 50{,}000$ from its location at one of the Nasmyth foci of the Spanish Gran Telescopio Canarias, still the largest optical-infrared telescope in the world with its 10.4 m segmented mirror; it pulled out of the race. Another project was a USA-UK Precision Radial Velocity Spectrometer for the Gemini North telescope (Ramsey et al. 2007), which later reincarnated in the UKIRT Planet Finder (Jones et al. 2013) and the Habitable Planet Finder for the Hobby-Eberly Telescope (Mahadevan et al. 2012). The Japanese proposed a state-of-the-art spectrograph for the Subaru telescope, namely InfraRed Doppler instrument, which can be fed by the telescope's adaptive optics system (Tamura et al. 2012). But the first of all was GIANO (Oliva et al. 2004), which was proposed to be an infrared (0.9-2.5 μm) cross-dispersed échelle spectrograph for the Italian Telescopio Nazionale Galileo. All these pioneers led the way to many others that came afterwards, such as SPIRou, CRIRES+, MAROON-X, NIRPS, Veloce and, especially, CARMENES, which was the first of them to see first light and to operate regularly.

CARMENES *[kár-men-es]* (Calar Alto high-Resolution search for M dwarfs with Exoearths with Near-infrared and optical Échelle Spectrographs) is the current last-generation instrument designed and built for the 3.5 m telescope at the Calar Alto Observatory by a consortium of German and Spanish institutions (Quirrenbach et al. 2020). It consists of two separated spectrograph channels covering the wavelength ranges from 0.52 μm to 0.96 μm and from 0.96 μm to 1.71 μm with spectral resolutions $R = 80{,}000\text{-}100{,}000$, each of which performs high-accuracy RV measurements (~1 m/s) with long-term stability. The fundamental science objective of CARMENES has been to carry out a survey of ~300 late-type main-sequence stars with the goal of detecting Earth-mass planets in their habitable zones and, incidentally, learn how spectra of M dwarfs actually look like. This chapter summarises how the homonymous CARMENES consortium made "their wildest dreams come true".

## CARMENES: the consortium

CARMENES is the name of the instrument (Quirrenbach et al. 2013, 2014), the consortium that designed and built it (Quirrenbach et al. 2010; Amado et al. 2013; García-Vargas et al.

2016), and of the main science project that has been carried out during guaranteed and legacy time observations since the start of operations (Reiners et al. 2018b; Ribas et al. 2023). Almost 300 scientists and engineers of 11 institutions in Spain and Germany, as well as a few dozen external guests, have participated in the design, construction and science exploitation of the new "planet hunter". The consortium was built on the basis of a membership and funding parity, with the same number of involved institutions in each country: five in Spain, five in Germany and the Spanish-German observatory of Calar Alto. The 11 institutions are enumerated in Table 1.

**Table 1** Institutions in the CARMENES consortium.

| Acronym | Institution | Head office | City, country |
|---------|-------------|-------------|---------------|
| MPIA | Max-Planck-Institut für Astronomie | Max-Planck-Gesellschaft | Heidelberg, Germany |
| IAA | Instituto de Astrofísica de Andalucía | Consejo Superior de Investigaciones Científicas | Granada, Spain |
| LSW | Landessternwarte Königstuhl | Zentrum für Astronomie der Universität Heidelberg | Heidelberg, Germany |
| ICE | Institut de Ciències de l'Espai | Consejo Superior de Investigaciones Científicas | Barcelona, Spain |
| IAG[a] | Institut für Astrophysik und Geophysik | Georg-August-Universität Göttingen | Göttingen, Germany |
| UCM | Universidad Complutense de Madrid | | Madrid, Spain |
| TLS | Thüringer Landessternwarte Tautenburg | | Tautenburg, Germany |
| IAC | Instituto de Astrofísica de Canarias | | La Laguna, Spain |
| HS | Hamburger Sternwarte | Universität Hamburg | Hamburg, Germany |
| CAB | Centro de Astrobiología | Consejo Superior de Investigaciones Científicas, Instituto Nacional de Técnica Aeroespacial | Madrid, Spain |
| CAHA | Centro Astronómico Hispano en Andalucía | Consejo Superior de Investigaciones Científicas, Junta de Andalucía | Almería, Spain |

Notes. [a] IAG was the Institut für Astrophysik Göttingen.

The Calar Alto observatory is located near Almería in southern Spain. It was operated jointly by the German Max-Planck-Gesellschaft and the Spanish Consejo Superior de Investigaciones Científicas (CSIC) until 2018; afterwards, the observatory of Calar Alto became fully Spanish. It hosts the 3.5 m, 2.2 m, 1.23 m and Schmidt telescopes, which started operating between the late 70s and mid 80s, as well as a number of other telescopes of various ownerships and diameters. In particular, the 3.5 m Zeiss Calar Alto telescope is the largest optical-infrared astronomical facility in mainland western Europe (Fig. 1).

The joint German-Spanish CARMENES consortium was a consequence of the merge of several letters of intent presented to the call, issued by the Calar Alto observatory, for the next instrument for the 3.5 m telescope in March 2008. After a public competition in early summer and an Observatory's decision just by the end of the same year, the CARMENES project started in February 2009 in Heidelberg, Germany, with a kick-off meeting for placing the stones for a conceptual design review eight months later.

The CARMENES Core Management Team was in charge of the day-to-day administration during the design, assembly-integration-verification and commissioning phases. This team included the principal investigators (PIs), project scientists (PSs), system engineers (SEs), project managers (PMs) and the observatory liaison. There was also a country parity in charges (e.g., German PI, Spanish co-PI; Spanish PS, German co-PS — The PM, co-PM, SE and co-SE rotated between Germany and Spain).

The consortium has been a focus of high-tech industrial development, since tens of contracts with industrial collaborators and manufacturers have been issued for completing the instrument, and some of the CARMENES designs have been reused by other teams worldwide. Although some key hardware had to be bought in the USA or Japan (e.g., near-infrared detector array, échelle gratings, photomultipliers), many components were purchased in Europe in general and Spain and Germany in particular. Some important providers in the two countries, in term of technical relevance and contract amount, were Ecofred, Fractal, Kaufman, Kinkele, NTG, Pfeiffer, Trinos, CeramOptec, Cryotherm, Cryovac, EHD, ESB DSO, Glenair, Huber, ILFA, Indicam, NC Solutions, Maassen, NTE-SENER, Proactiverd, Qucam, Vacom, and Xu OSE. The consortium also collaborated with ESO on designs of detector cryostats, sorption pumps and the near-infrared channel cooling unit. Further collaborations with high-technology companies in Europe are planned for the near future.

During the CARMENES operations phase, which started in January 2016, the most active consortium body has instead been the Science Coordination Team, which includes representatives of the 11 institutions in Table 1. Hence, during the first year of operations the CARMENES consortium smoothly evolved from an engineering-oriented project to a science-oriented one in which young postdocs and students at all levels play a role. To date, 33 PhD, 53 MSc, and 17 BSc theses directly related to CARMENES have been defended, and many others are ongoing. Such theses cover a wide range of topics on instrumentation, M-dwarf stars, and exoplanets.

Far from the cliché, the German-Spanish collaboration (Fig. 2) has been extremely successful in spite of the 2007-2008 financial crisis, which impacted on the European economies for several years, and other embarrassing institutional matters of severe importance that cannot be described in this chapter. One way of quantifying that extreme successfulness is with a pair of example facts: (1) in the CARMENES final design review documentation presented in February 2013, the CARMENES end of commissioning of the two spectrograph channels was scheduled for December 2015. On the contrary of what is usual in complex instrument projects, it was accomplished one month *earlier*; and (2) in February 2023, exactly ten years after the final design review, the CARMENES consortium published its 100th refereed paper.

## CARMENES: the instrument

CARMENES is installed at the 3.5 m Zeiss Teleskop ("*el tres y medio*") at the Calar Observatory, at 2168 m in Gérgal, in the Aldalusian province of Almería in the southeast of Spain. CARMENES has many technologies in common with other "extreme precision spectrographs", but also others that make them unique. As well as HARPS and many other planet hunters, CARMENES is a white-pupil, fibre-fed, grism cross-dispersed, échelle spectrograph working in quasi-Littrow mode (Fig. 3). CARMENES is, however, a dual-channel spectrograph. On the contrary to ESPRESSO at the Very Large Telescope, the most advanced spectrograph to date, the two channels of CARMENES are enclosed in two independent, almost identical, vacuum tanks with their respective optical benches. There is a dichroic at 0.96 μm that splits the light into two beams, one for each channel, namely VIS (from "visible") and NIR (from "near infrared"). The detectors are actually sensitive until 1.05 μm (a CCD in VIS) and from 0.90 μm (a CMOS array in NIR). There are some gaps in the wavelength coverage wider than 0.01 μm redwards of 1.55 μm, while the strongest telluric absorption is at the boundary between the *J* and *H* photometric bands (Fig. 4). The most important parameters of the CARMENES instrument are summarised in Table 2.

The two VIS and NIR tanks are in one isolated climatic room each, with a third climatic room for the two calibration units. The technical corridor that connects the three rooms contains also racks for the interlocks system, detector-dedicated computers and other electronics, as well as the large nitrogen bottles for both VIS and NIR detector cryostats and the NIR channel cooling unit, as described below.

**Table 2** Summary of the CARMENES instrument.

| Parameter | VIS channel | NIR channel |
|---|---|---|
| Wavelength coverage | 0.52-0.96 μm ([*V*]*RIZ*) | 0.96-1.71 μm (*YJH*) |
| Detector | 1 × 4k×4k e2v CCD231-84 | 2 × 2k×2k Hawaii-2RG (2.5 μm cutoff) |
| Wavelength calibration | Th-Ne, U-Ne, U-Ar hollow cathode lamps, Fabry-Pérot etalon | |
| Working temperature | 284 K | 138 K |
| Spectral resolution | 94,600 | 80,400 |
| Mean sampling | 2.8 pixels | |
| Mean inter-fibre spacing | 7.0 pixels | |
| Cross disperser | Grism, LF5 glass | Grism, infrasil |
| Reflective optics coating | Silver | Gold |
| Number of orders | 55 | 28 |
| Échelle grating mosaic | 2 × Richardson Gratings R4 (31.6 mm$^{-1}$, 76 deg) | |
| Target fibre field of view | 1.5 arcsec | |
| A&G[a] system field of view | 3 arcmin | |
| A&G[a] system band | Approx. *R* | |

Notes. [a] A&G stands for acquisition and guiding.

While the two spectrograph channels with their enclosures and auxiliary devices are in the old coudé room of the telescope, just one level above it, in the dome, there is a key CARMENES component: the front-end. The photons from an on-sky target suffer only two reflections on the primary and secondary mirrors of the 3.5 m telescope before reaching the CARMENES front-end at the telescope's Cassegrain focus (Fig. 5). The CARMENES front-end is not only the instrument component that transmits the light from a sky target and the calibration units to the spectrographs themselves, but also the physical interface with the telescope (above it) and the other instrument currently at the 3.5 m telescope's Cassegrain focus (below it): the Potsdam Multi-Aperture Spectrophotometer (PMAS; Roth et al. 2005). PMAS is a several-ton integral field spectrophotometer that covers from 0.35 μm to 0.90 μm in low spectral resolution, which hangs from the CARMENES front-end. The first optical surface of CARMENES is, therefore, a movable pick-up mirror that intersects the light beam from the telescope and allows to switch in less than 2 min between CARMENES and PMAS or the instrument that will eventually replace it, the Tetra-ARmed Super-Ifu Spectrograph (TARSIS; Gil de Paz et al. 2020). In the few centimetres left between the massive telescope primary mirror support and PMAS/TARSIS (24.7 cm exactly), the steel-and-aluminum 220 kg-mass CARMENES front-end contains a number of opto-mechanical components. Sorted by order of light reflection/transmission, besides the motorised, 45-deg pick-up mirror on rails, there are a two-double-prism atmospheric dispersion corrector, the dichroic beam splitter, an acquisition and guiding system, and the VIS and NIR fibre heads. Since the dichroic reflects the VIS light and transmits the NIR light, there is an additional NIR mirror. The VIS fibre head, which mount has besides a shutter for the CCD, is located behind a small hole in a mirror towards the guiding camera. Acquisition is made easy thanks to the extraordinary pointing accuracy of the 3.5 m telescope, of a couple of arcseconds, and an LED that illuminates the mirror and its hole before exposing. Guiding is performed on the wings of the target star reflected by the mirror (on-axis), on the stars in the 3 arcmin-field of view (off-axis) and, in most occasions, in a hybrid mode (both on- and off-axis with weights depending on the seeing, automatically-selected exposure time, and $R$-band magnitudes of the primary targets and of other stars in the field of view). The target fibre (i.e., the hole in the VIS fibre head mirror) projected on the sky has a diametre of 1.5 arcsec, which is a compromise between a small light beam size at the spectrograph entrance and an aperture large enough to enclose the median seeing at Calar Alto and to minimise losses under bad weather conditions. There are actually two holes separated by 88 arcsec in the sky, the one for the target spectrum to the east, the other for the sky background spectrum to the west if no simultaneous wavelength calibration is selected. Additionally, in front of each of the fibre entrances, there is a movable, finger-like, flat, mirror that acts as the observing mode selector (see below).

The front-end is connected to the coudé room via four optical fibres: the input for the VIS and NIR channels and the output of the VIS and NIR calibration units. Actually, there are eight fibres because all of them are redundant in case of accident. The first 60-80 m of the channel fibres are circular and with a 100 μm core diametre, but the last 2-3 m after a double ferrule junction (FC/FC) are octagonal, for maximising the scrambling effect of the fibres when the guiding is not optimal. Besides, for suppressing modal structures, the last segment of the on-sky circular is agitated by a fibre shaker, which basically bends the fibre in two alternate perpendicular directions (the fibre shakers of both channels were turned off after five years of operations, without an impact on the RV precision). The on-sky octagonal fibre enters the corresponding spectrograph channel vacuum tank through a feed through. It consists of a plate with two FC receptacles where the space between the two connectors was filled and sealed with a low-viscosity epoxy resin. The fibres from the calibration units to the front-end are, however, only circular and thicker (600 μm core). All the fibres are protected with a kevlar and polyvinyl chloride tubing and inserted in pairs in plastic hosepipes.

The two fibres entering each spectrograph channel can be illuminated with different light sources. This is selected by extending and retracting the observing mode-selector mirror in the front-end, which can reflect the light from none, one or two fibres from the respective calibration unit. In short:
- Fully extended mirror — in front of both calibration fibre beams: for daily calibration operations (i.e., the same halogen or hollow-cathode lamp or Fabry-Pérot etalon in both fibres).
- Half extended mirror — in front of only the lower calibration fibre beam: for standard

nightly observations during the main programme (i.e., light from star in one fibre, light from the Fabry-Pérot etalon in the other fibre).
- Fully retracted mirror — out of beams: for other programmes that do not require precise wavelength calibration (i.e., light from star in one fibre, light from the sky background in the other fibre).

During the design process, the instrument team tried to keep as much commonality between the two spectrograph channels as possible. Each of the almost identical vacuum tanks that enclose the spectrograph channels consists of a cylindrical section of 1.5 m in diameter and 3.0 m in length to fit the 1600 kg-mass optical bench, plus two asymmetrical end caps (Fig. 6). The cold-rolled stainless steel wall thickness was set to 12 mm. The optical mounts and radiation shields were assembled on the optical benches, which were moved in and out of each vacuum tank on a rail system during the assembly, integration and verification phase. Each full tank is supported on four active pneumatic insulators for damping vibrations (although the telescope support and the coudé room are physically separated parts of the building). These damping legs are connected to the dry air circuit of the telescope. Apart from a large 500-mm flange in the rear cap for the detector cryostat, a number of smaller DN (diameter nominal) flanges and feed throughs in the main cylinder are necessary for the vacuum pumps, temperature sensors (up to 24), detector cooling, cryostat and electronics, cooling gas in- and outflow (only in the NIR channel) and, of course, the optical fibres from the front-end. With the current vacuum system, the actual pressure inside both tanks is almost two orders of magnitude lower than the engineering requirements at $10^{-3}$ mbar.

Once inside each of the vacuum tanks, the light exiting from the optical fibres through microlenses is converted to the collimator focal ratio with an F/N system and sliced into two beams by a two-slice image slicer of the Bowen-Walraven type. This slicing is done, in its turn, on the two beams for the target spectrum and for the simultaneous calibration or background sky spectrum. This unit, the first one on the optical bench that the photons go through, is rotated to align the two pairs of half-moon slices with the detector pixel arrangement. Next, the light reflects on the three-pass collimator mirror (made of a single zerodur paraboloid), the R4 échelle grating mosaic that produces the main dispersion, the collimator again, a flat mirror, and the collimator for the third and last time, and then refracts through a grism cross-disperser that separates the orders and a fully dioptric six-lens camera (Fig. 7). In the narrow space between the last camera lens and the detector head dewar there is a cylindrical field lens, which is the last surface before the photons hit the detector. Several field stops are also mounted on each optical bench for baffling stray light.

In every spectrograph channel, the zeroth-order light from the échelle grating is collected by an off-axis parabolic mirror behind and above the collimator, and routed outside the vacuum tank via a 1000 μm-core fibre to a photo-counter unit. With this system, the instrument records the received intensity with high time resolution (1 s) and precisely measures the photon-weighted midpoint of each exposure. This is needed for an accurate conversion of the observed radial velocity to the barycentre of the Solar System. The instrument also uses the running photon counts to make real-time adjustments to the integration time depending on atmospheric conditions. The instrument team called this system the exposure-meter.

Each channel has a calibration unit, which is installed in the calibration room of the coudé area and is connected to the front-end with a 400 µm fibre link. CARMENES uses Th-Ne, U-Ne and U-Ar hollow-cathode emission line lamps for absolute wavelength calibration and tungsten lamps for flat fielding. The different lamps (the lamp holders have four different colours, one for each type) in each calibration unit are installed in an octagonal arrangement, dubbed the lighthouse. The light of any lamp in the octagonal lighthouse is directed to the calibration unit optical bench with a rotating 45-deg mirror. On the optical bench there is a 50/50 beam splitter that divides the beam into two, which go to the front-end through the pair of 400 µm optical fibres. The beam splitter can also inject a second light source into one of the fibres (Fig. 8). In the CARMENES case, close to both calibration units there are two Fabry-Pérot etalons, each connected via fibre to its respective calibration unit. This etalons are calibrated against the hollow-cathode lamps during automatic calibration runs just before sunset and just after sunrise. As detailed below, the original etalons were completely replaced in Spring 2024.

The CARMENES standard observing mode for their planet survey is injecting light from the Fabry-Pérot etalon only in the calibration optical fibre. Since hollow-cathode lamps that are used nightly show ageing effects, they are monitored regularly through comparisons against older spectra and master lamps that are used either once per day or only occasionally, and ultimately against super-master lamps that ensure long-term stability. The super-master lamps are stored in a dedicated tank, also in the calibration room, filled with a low-pressure Ne atmosphere.

In spite of the commonality in the design of the two spectrographs, there are two fundamental differences between the VIS and NIR channels because of the distinct wavelength range: the detectors and the operational temperature. They both led to a cascade of additional minor differences. First, the VIS channel is equipped with a 4k × 4k deep-depleted, back-illuminated e2v detector CCD231-84 with suppressed fringing and graded coating, such that the efficiency is maximum for each position in cross-dispersion direction. Some mechanical and electronics solutions in the VIS channel detector unit (i.e., preamplificators, controller, power supply, rigiflex printed circuit board, continuous-flow cryostat) had been already used by ESO in the Multi Unit Spectroscopic Explorer (MUSE) integral-field spectrograph. Actually, the mechanics of the VIS detector head is a legitimate copy of the MUSE one. On the other hand, the NIR channel is equipped with a mosaic of two 2k × 2k Teledyne Hawaii-2RG substrate-removed detectors with the standard 2.5 µm cutoff and a custom-made read-out electronics (dubbed "Tom" and "Jerry"). Again, the instrument team used their experience with similar instrumentation (i.e., LUCI-2 for the Large Binocular Telescope, PANIC for Calar Alto) in the design, construction and characterisation of the NIR detector unit.

Second, both VIS and NIR channels have thermally isolating mechanical links (from the optical components to the mechanical mounts, the optical bench and three supports to the walls of the vacuum tanks). Thanks to the combination of insulation and thermal inertia of the massive optical benches, the temperature sensors display milli-Kelvin variations in scales of weeks ($\sigma(T)$ = 4–7 $10^{-3}$ K over 30 d). However, while the VIS channel operates at (climatised) room temperature, at about +11 ºC, the NIR one operates at a much cooler temperature, at about -135 ºC (see Table 2 for the values in K). For that, the NIR channel has its own cooling unit and 20-layer insulation blankets as radiation shields surrounding the optical bench. The NIR channel cooling unit is based on the semi-continuous flow of nitrogen gas at very low temperature. This gas is injected into a pipe circuit attached to the inner and outer parts of the radiation shield. There is also a circuit extension attached to the optical bench itself, but it was used only during the the instrument pre-cooling before operations. The key component of the NIR cooling system is the $N_2$ gas preparation unit (N2GPU), which basically consists of a set of intermediate and final heat exchangers and an evaporator unit connected to liquid $N_2$ inlet and gas $N_2$ outlet tubes inside a stainless steel vessel. The unit is about 1 m wide not counting all the corresponding pumps and controllers, and can be considered as a prototype for the large detector cryostats that will be necessary at the Extremely Large Telescope, now under construction. Because of the N2GPU and the large liquid $N_2$ bottles to fill the VIS and NIR detector cryostats and the N2GPU itself, the instrument team also installed $O_2$ sensors and alarms in the coudé area for safety reasons. Since summer 2023, the N2GPU has a continuous (instead of discontinuous) flow of $N_2$ gas, which has helped in improving its thermal stability (see also below).

CARMENES has a completely fixed spectrograph configuration. Apart from the pumps, which have operated only at certain times (e.g., after sorption pump regeneration), the only movable parts of the instrument are:
- In the front-end, the 45 deg pick-up mirror, the two lenses of the atmospheric dispersion corrector, and the VIS and NIR mode-selector finger mirrors.
- Also in the front-end, the shutter of the VIS channel.
- In the coudé area, inside each climatic room, the VIS and NIR fibre shakers (now turned off).
- In each of the VIS and NIR calibration units, the rotating mirror inside the octagon, high-velocity (millisecond) shutters at the entrance of the calibration fibre and at the output of the Fabry-Pérot fibre (there is also a rotating wheel with notch filters, but which is always open).

All this complex machinery is controlled by the instrument control system, which supervises and manages all the sensors and subsystems of the instrument, and coordinates the taking of exposures with the telescope control system. For regular survey operations, CARMENES follows a well structured data flow that we summarise below (Fig. 9):

1. The CA[RMENES] Scheduling Tool (`CAST`; García Piquer et al. 2017) takes into account observational constraints and distributes the awarded telescope time with genetic algorithms amongst the targets of the GTO, legacy and CARMENES-TESS programmes. Previously, the basic parameters of all the to-be-observed M dwarfs (and selected K dwarfs), especially equatorial coordinates and magnitudes, are compiled by the CARMENES input catalogue (Carmencita; Caballero et al. 2016a). Since all the observations are being executed in service mode by observatory staff, GTO and legacy targets are accompanied by customised finding charts that took into account the high proper motion of the stars.

2. A few seconds after semi-automatic acquisition and start of guiding, the VIS and NIR channels observe simultaneously until a signal-to-noise ratio S/N = 150 at order 50 (~1.2 µm) on the NIR spectrum is attained or a maximum time of 1800 s is exposed, whatever occurs first. Since the VIS CCD takes approximately 33 s to be read, the NIR CMOS meanwhile goes on exposing for an extra half minute. The S/N of the NIR spectrum is derived from a calibrated relationship with the number of counts on the NIR exposure-meter, although the system can also use VIS as the master channel (for early-type stars). The maximum exposure time was set to 1800 s to avoid excessive line broadening due to Earth's rotation and contamination from cosmic rays.

3. Every target visit provides the same set of raw data: one or several acquisition images (FITS), the VIS and NIR spectra (FITS) and the VIS and NIR exposure-meter temporal series (ASCII). The two raw spectra are stored in heavy FITS files with comprehensive headers (programme, target, environmental conditions, dome and telescope, all instrument units, time at median point of exposure, etc.).

4. As in all CARMENES observation programmes, regardless they are executed under guaranteed, legacy or open time, the raw data go through the first pipeline, namely `CARACAL` (CA[RMENES] Reduction And CAlibration; Zechmeister et al. 2014). `CARACAL` automatically makes the dark and bias corrections, order tracing, flat-relative optimal extraction and wavelength calibration, generates four fully reduced one-dimensional spectra (two channels × two fibres) and computes a rough estimation of the absolute radial velocity of the target by comparison with a mid-M dwarf theoretical model after barycentric correction. Usually, and in spite of the relative speed of the 3.5 m telescope for re-pointing, these four spectra are generated and automatically displayed by the graphical user interface in the control room before the next target is acquired. All these data are stored in real time at the observatory repository.

5. Next morning, the consortium's data at the observatory repository are synchronised with the CARMENES Data Archive in Madrid, which makes all the spectroscopic data available to the rest of the consortium through a password-protected website. A local mirror of the archive is also in Göttingen. The reduced spectra are next processed through another two pipelines, one that makes the telluric absorption correction on the targets' spectra with the template division telluric modelling technique (TDTM; Nagel et al. 2023) and other one that computes RVs, spectral activity indicators and high-S/N stacked spectra, namely "template spectra". The later pipeline, the SpEctrum Radial Velocity AnaLyser (`SERVAL`; Zechmeister et al. 2018), is being extensively used by CARMENES and other teams worldwide. `SERVAL` is complemented by another open source python package, Radial velocity and Activity indicators from Cross-COrrelatiON with masks (`RACCOON`; Lafarga et al. 2020). Finally, all the `SERVAL` and `RACOON` output files for each target are stored in the Science Wiki, from which the CARMENES consortium members retrieve them for preparing their publications. Additional codes have been implemented for the scientific exploitation of the reduced CARMENES data, such as `StePar` (Tabernero et al. 2019) and `SteParSyn` (Tabernero et al. 2022) to infer stellar atmospheric parameters.

In total, there are eleven computers controlling CARMENES: *gliese* and *jahreiss* for the instrument control system itself, *wolf* and *ross* for the VIS and NIR detectors, *barnard* and *lalande* for the VIS and NIR pipelines, *kapteyn* and *lacaille* for the VIS and NIR temporary data storage, *giclas* for the acquistion and guiding camera, *struve* for the interlocks and *luyten* for the graphical user interface (some of them have clones, mirrors or back-ups: *ross*, *wolf*, *giclas*, *struve*). Besides, a twelfth computer, namely *vb* (from van Biesbroeck), stores the CARMENES Data Archive.

Further details on CARMENES are provided in a long series of SPIE publications on a number of different topics: NIR cooling unit (Becerril et al. 2010, 2012, 2016a, 2016b; Amado et al. 2012; Mirabet et al. 2014), instrument control system, scheduling and interlocks (Guàrdia et al. 2012; García Piquer et al. 2014; Colomé et al. 2016; Helming et al. 2016), spectrograph and front-end opto-mechanics (Seifert et al. 2012, 2016; Tala-Pinto et al. 2018), calibration units (Sarmiento et al. 2014; Schäfer et al. 2018), optical fibres (Stürmer et al. 2014), data flow (Caballero et al. 2016b), system engineering (Pérez-Calpena et al. 2016) and project management (García-Vargas et al. 2016). Besides, every two years since the beginning of the project until the end of the guaranteed time observations the CARMENES consortium summarised the latest developments and results in special all-hand issues (Quirrenbach et al. 2010, 2012, 2014, 2016, 2018, 2020, 2024).

On-site operations have been and are currently performed by the observatory staff, while the instrument team still provides services such as automated scheduling, monitoring of instrument health and data quality, and pipeline processing of all data. Joint efforts have been necessary to implement measures to improve the performance, and to address occasional problems and failures (Quirrenbach et al. 2024). As a consequence of the COVID-19 pandemic, observations at CAHA had to be suspended on 16 March 2020. Only a minimal presence of CAHA staff on the mountain could be maintained to perform essential maintenance such as surveillance of equipment, servers, and internet connection, to provide LN2 support, and to keep CARMENES in a stable state allowing a quick transition into an operational mode. Because of that, and because of the autonomy and remoteness of the observatory, it was possible to resume data taking with CARMENES already on 6 May 2020, so the loss of observing time was relatively minor.

As a consequence of the spectral energy distribution of M dwarfs, along with the density and depth of their spectral features in different wavelength regions, the photon-limited information content of the VIS spectra was higher than that of NIR for almost all stars in the original CARMENES GTO survey sample (Reiners et al. 2018b; Reiners & Zechmeister 2020). Therefore, the VIS RVs have almost always smaller internal errors than those obtained simultaneously in the NIR, which justifies the practice in many publications from CARMENES to use only the RVs from the VIS channel. According to Ribas et al. (2023), the GTO survey achieved a median internal precision of 1.27 m/s for 362 target stars, while the observed RV rms for 344 of these stars (excluding binaries) had a median value of 3.82 m/s. From these values, one can conclude that for most stars the RV variations are dominated by astrophysical "jitter" (i.e., stellar variability) and the signals from planets themselves rather than residual instrumental effects.

**Table 3** Scheduled hours and number of observations carried out during GTO and Legacy+ projects.

| Project | Year | Scheduled time (h) | $N_{\mathrm{obs}}$ |
|---|---|---|---|
| GTO | 2016 | 994 | 4269 |
| | 2017 | 1185 | 4539 |
| | 2018 | 1064 | 3517 |
| | 2019 | 1216 | 3721 |
| | 2020 | 964 | 2719 |
| Legacy+ | 2021 | 451 | 1763 |
| | 2022 | 424 | 1735 |
| | 2023 | 533 | 2396 |

Table 3 lists the number of hours scheduled for the GTO and legacy (dubbed Legacy+) programs, not including weather losses, and the number of observations for each year. The step from 2020 to 2021 reflects the transition from GTO to Legacy+. Other variations in the number of spectra from year to year are largely due to different mixes of relatively bright and faint stars. In general, the average integration time has increased over time, as the main survey has reached the number of desired spectra preferentially for bright stars, and is now progressing towards the faintest targets in the sample (Quirrenbach et al. 2024).

Additional improvements to CARMENES are being presently carried out, which are encompassed as part of a new project dubbed CARMENES+. Most of the CARMENES+ activities between 2020 and 2023 focused on changing the operation mode of the NIR N2GPU from discontinuous to continuous (plus a plethora of minor modifications to the NIR cooling system; R. Calvo-Ortega et al. in prep.), as well as on modifying the daily procedure of exchanging the large empty and filled $N_2$ bottles of the N2GPU in the technical corridor, which previously was completely manual. Since 2024, the upgrade activities have focussed instead on the wavelength calibration system. First, in April 2024 the engineering team replaced the two tanks of the VIS and NIR Fabry-Pérot etalons by a single one that covers with a finesse better than 80% in the whole range from 0.52 μm to 1.71 μm (already done). The new cryostat takes also advantage of better insulation and light injection, as well as new pressure and temperature inner sensors (Fig. 10). The team also switched the operation mode to keep the vacuum pumps of the VIS and NIR main tanks running 24/7. Next, the engineering team plans to exchange the optical fibre injection to the calibration units with customised fibre switches that will allow us to simultaneously record the spectra of hollow cathode lamps and the new Fabry-Pérot etalons (S. Schäfer et al. in prep.). These upgrades must be done in conjunction with minimal modifications to the instrument control system.

At this final stage, and only for illustrative purposes, CARMENES might be compared to other ultra-precise spectrographs. Being at a telescope of the same size, it has a lower spectral resolution than the HARPS+NIRPS combination at the 3.6 m ESO telescope on La Silla. Likewise, CARMENES covers neither the Ca II H&K doublet nor the *K* band as HARPS+NIRPS does, but instead does not have a wavelength gap between 0.69 μm and 0.95 μm, where the Ca II infrared triplet and, especially, most of the RV information in early and mid M dwarfs resides. The comparison with GIARPS (HARPS-N+GIANO-B) at the Telescopio Nazionale Galileo is identical, except for the fact that GIANO-B has a lower spectral resolution than the NIR CARMENES channel. ESPRESSO and MAROON-X at 8 m-class telescopes have much better short-term RV precision than CARMENES, but do not cover the near infrared and, being at telescopes with large oversubscription factors, are unable to carry out large long-term RV programmes with satisfactory sampling. There are other excellent infrared high-resolution spectrographs worldwide, such as IRD, HPF and SPIRou, but none of them has an optical channel. The comparison could go on and on with iSHELL, Veloce, Keck's HIRES and many others. However, in science in general and in exoplanet surveys in particular, there should be no "best" or "worst", but all facilities should complement the others, and the different teams should work elbow to elbow. For that reason, as the reader will see below, some of the best CARMENES scientific results have been obtained in collaboration with other teams outside the German-Spanish consortium.

## CARMENES: the science

Long story short, CARMENES provides one of the best sensitivities for the detection of low-mass planets and the tools for discriminating between genuine planet detections and false positives caused by stellar activity, specially in M dwarfs. Besides, it is a well-oiled machine that has robustly operated without major flaws since 2016, and among their ranks there are first-class researchers with a wide expertise in a number of topics, from stellar parameters and activity, through photometric variability and exoplanet atmospheres, to radial velocity analysis. For that reason, CARMENES has assembled a noteworthy publication record that can only be briefly summarised here. Below we show just three topical examples.

**Table 4** Planets discovered by CARMENES or with CARMENES contribution.

| Star | GJ or TOI | $d$ [pc] | Sp. type | Planet | $P$ [d] | $M_2$ [$M_\oplus$] | Reference [a] |
|---|---|---|---|---|---|---|---|
| GX And | 15 A | 3.562 | M1.0 V | c | 6694 | > 50.4 | Tri18 |
| Ross 1003 | 1148 | 11.03 | M4.0 V | c | 532.6 | > 72.1 | Tri18 |
| HD 147379 | 617 A | 10.77 | M0.0 V | b | 86.57 | > 23.0 | Rei18a, Rib23 |
| HD 180617 | 752 A | 5.915 | M2.5 V | b | 105.90 | > 12.2 | Kam18 |
| LP 819-52 | 1265 | 10.24 | M4.5 V | b | 3.6511 | > 7.4 | Luq18 |
| Ross 1020 | 3779 | 13.75 | M4.0 V | b | 3.0232 | > 8.0 | Luq18 |
| G 232-70 | 4276 | 21.30 | M4.0V | b | 13.352 | > 16.57 | Nag19 |
| HD 119130 | … | 114.3 | G3 V | b | 16.9841 | 24 | Luq19a |
| K2-285 | … | 156 | K2 V | b | 3.43175 | 9.7 | Pal19 |
| | | | | c | 7.13804 | 16 | Pal19 |
| | | | | d | 10.4560 | < 6.5 | Pal19 |
| | | | | e | 14.7634 | < 10.7 | Pal19 |
| BD+61 195 | 49 | 9.860 | M1.5 V | b | 13.8508 | > 5.63 | Per19 |
| Teegarden's | … | 3.832 | M7.0 V | b | 4.90634 | > 1.16 | Zec19, Dre24 |
| | | | | c | 11.416 | > 1.05 | Zec19, Dre24 |
| | | | | d | 26.13 | > 0.82 | Dre24 |
| LSPM J2116+0234 | … | 17.64 | M3.0 V | b | 14.4433 | > 13.4 | Lal19 |
| GJ 357 | 357 | 9.444 | M2.5 V | b | 3.93072 | 1.84 | Luq19b |
| | | | | c | 9.1247 | > 3.40 | Luq19b |
| | | | | d | 55.661 | > 6.1 | Luq19b |
| LP 90-18 | 3512 | 9.489 | M5.5 V | b | 203.13 | > 146 | Mor19, Rib23 |
| | | | | c | 2350 | > 143 | Mor19, Rib23 |
| TYC 6170-95-1 | … | 323.1 | G8 IV/V | b | 3.5949 | 8.8 | Hid20 |
| | | | | c | 15.624 | 14.7 | Hid20 |
| | | | | d | 35.747 | 10 | Hid20 |
| HD 79211 | 338 B | 6.334 | M0.0 V | b | 24.45 | > 10.27 | Gon20 |
| CD Cet | 1057 | 8.069 | M5.0 V | b | 2.29070 | > 3.95 | Bau20 |
| TOI-1235 | TOI-1235 | 39.68 | M0.5 V | b | 3.444717 | 5.90 | Blu20 |
| LP 729-54 | TOI-732 | 22.0 | M3.5 V | b | 0.76837931 | 2.46 | Now20, Bon23 |
| | | | | c | 12.252284 | 8.04 | Now20, Bon23 |
| G 50-16 A | 3473 | 27.32 | M4.0 V | b | 1.98003 | 1.86 | Kem20 |
| | | | | c | 15.509 | > 7.41 | Kem20 |
| HD 238090 | 458 A | 15.25 | M0.0 V | b | 13.671 | > 6.89 | Sto20b |
| HD 265866 | 251 | 5.585 | M3.0 V | b | 14.2383 | > 4.27 | Sto20b, Rib23 |
| LP 714-47 | TOI-442 | 52.34 | M0.0 V | b | 4.052037 | 30.8 | Dre20 |
| Wolf 437 | 486 | 8.083 | M3.5 V | b | 1.4671205 | 3.00 | Tri21, Cab22 |
| GJ 740 | 740 | 11.11 | M1.0 V | b | 2.37756 | > 2.96 | Tol21 |
| LP 14-53 | TOI-1640 | 18.21 | M3.0 V | b | 1.949538 | 2.33 | Sot21 |
| TOI-1685 | TOI-1685 | 37.61 | M3.0 V | b | 0.669140 | 3.76 | Blu21 |
| | | | | c | 9.02 | > 9.2 | Blu21 |
| G 264-12 | … | 15.99 | M4.0 V | b | 2.30538 | > 2.50 | Ama21 |
| | | | | c | 8.0518 | > 3.75 | Ama21 |
| BD+01 2447 | 393 | 7.038 | M2.0V | b | 7.02679 | > 1.71 | Ama21 |
| TOI-1238 | TOI-1238 | 70.64 | M0.0 V | b | 0.764597 | 3.8 | Gon22 |
| | | | | c | 3.294736 | 8.3 | Gon22 |
| TOI-1201 | TOI-1201 | 37.64 | M2.0 V | b | 2.491986 | 6.28 | Kos21 |
| TOI-1759 | TOI-1759 | 40.11 | M0.0 V | b | 18.85019 | 10.8 | Esp22 |
| V1298 Tau | … | 108.5 | K1 V | b | 24.1399 | 200 | Suá22 |
| | | | | c | 8.24892 | < 76 | Suá22 |
| | | | | d | 12.4058 | < 99 | Suá22 |
| | | | | e | 40.2 | 369 | Suá22 |
| G 180-18 | 3929 | 15.83 | M3.5 V | b | 2.616267 | 1.27 | Kem22 |
| G 161-32 | TOI-620 | 33.06 | M2.5 V | b | 5.0988179 | < 7.1 | Ree22 |
| TYC 2187-512-1 | … | 15.48 | M1.0 V | b | 691.6 | 105 | Qui22 |
| HD 260655 | 239 | 9.998 | M0.0 V | b | 2.76953 | 2.14 | Luq22 |
| | | | | c | 5.70588 | 3.09 | Luq22 |
| TOI-1468 | TOI-1468 | 24.71 | M1.5 V | b | 1.8805136 | 3.21 | Cha22 |
| | | | | c | 15.532482 | 6.64 | Cha22 |
| BD+11 2576 | 514 | 7.628 | M0.5 V | b | 140.43 | > 5.2 | Dam22 |
| GJ 1002 | 1002 | 4.846 | M5.5 V | b | 10.3465 | > 1.02 | Suá23 |
| | | | | c | 21.202 | > 1.36 | Suá23 |
| Wolf 1069 | 1523 | 9.575 | M5.0 V | b | 15.564 | > 1.26 | Kos23 |
| GJ 1151 | 1151 | 8.043 | M4.5 V | b | 389.7 | > 10.6 | Bla23 |
| HN Lib | 555 | 6.253 | M4.0 V | b | 36.116 | > 2.59 | Gon23a |
| BD+44 3567 | 806 | 12.06 | M1.5 V | b | 0.9263237 | 1.90 | Pal23 |
| | | | | c | 6.64064 | > 5.80 | Pal23 |
| TOI-1470 | TOI-1470 | 51.74 | M1.5 V | b | 2.527113 | 7.52 | Gon23b |
| | | | | c | 18.0881 | 7.94 | Gon23b |
| TOI-2095 | TOI-2095 | 41.90 | M2.5 V | b | 17.66484 | < 4.1 | Mur23 |
| | | | | c | 28.17232 | < 7.4 | Mur23 |
| BD-13 5069 | 724 | 16.99 | M1.0 V | b | 5.101284 | > 10.8 | Gor23 |
| G 203-42 | 3988 | 9.908 | M4.5 V | b | 6.9442 | > 3.69 | Gor23 |
| LP 375-23 | TOI-1801 | 30.89 | M0.5 V | b | 10.64387 | 5.74 | Mal23 |
| HD 110067 | TOI-1835 | 32.22 | K0 V | b | 9.113678 | 5.69 | Luq23 |

| | | | | c | 13.673694 | < 6.3 | Luq23 |
| | | | | d | 20.519617 | 8.5 | Luq23 |
| | | | | e | 30.793091 | < 3.9 | Luq23 |
| | | | | f | 41.05854 | 5.0 | Luq23 |
| | | | | g | 54.76992 | < 8.4 | Luq23 |
| HD 191939 | TOI-1339 | 53.48 | G9 V | g | 284 | > 13.5 | Ore23 |
| Wolf 327 | TOI-5747 | 28.56 | M2.5 V | b | 0.5734745 | 2.53 | Mur24 |
| G 182-34 | TOI-4438 | 30.06 | M3.5 V | b | 7.44628 | 5.1 | Gof24 |
| GJ 12 | TOI-6251 | 12.17 | M3.0 V | b | 12.761408 | < 3.9 | Kuz24 |
| GJ 4256 | TOI-6255 | 20.39 | M3.0 V | b | 0.23818244 | 1.44 | Dai24 |

Notes. [a] References — Ama21: Amado et al. 2021; Bau20: Bauer et al. 2020; Bla23: Blanco-Pozo et al. 2023; Blu20: Bluhm et al. 2020; Blu21: Bluhm et al. 2021; Bon23: Bonfanti et al. 2023; Cab22: Caballero et al. 2022; Cha22: Chaturvedi et al. 2022; Dai24: Dai et al. 2024; Dam22: Damasso et al. 2022; Dre20: Dreizler et al. 2020; Dre24: Dreizler et al. 2024; Esp22: Espinoza et al. 2022; Gof24: Goffo et al. 2024; Gon20: González-Álvarez et al. 2020; Gon22: González-Álvarez et al. 2022; Gon23a: González-Álvarez et al. 2023a; Gon23b: González-Álvarez et al. 2023b; Gor23: Gorrini et al. 2023; Hid20: Hidalgo et al. 2020; Kam18: Kaminski et al. 2018; Kem20: Kemmer et al. 2020; Kem22: Kemmer et al. 2022; Kos21: Kossakowski et al. 2021; Kos23: Kossakowski et al. 2023; Kuz24: Kuzuhara et al. 2024; Lal19: Lalitha et al. 2019; Luq18: Luque et al. 2018; Luq19a: Luque et al. 2019a; Luq19b: Luque et al. 2019b; Luq22: Luque et al. 2022; Luq23: Luque et al. 2023; Mal23: Mallorquín et al. 2023; Mor19: Morales et al. 2019; Mur23: Murgas et al. 2023; Mur24: Murgas et al. 2024; Nag19: Nagel et al. 2019; Now20: Nowak et al. 2020; Ore23: Orell-Miquel et al. 2023a; Pal19: Pallé et al. 2019; Pal23: Pallé et al. 2023; Per19: Perger et al. 2019; Qui22: Quirrenbach et al. 2022; Ree22: Reefe et al. 2022; Rei18a: Reiners et al. 2018a; Rib23: Ribas et al. 2023; Sot21: Soto et al. 2021; Sto20b: Stock et al. 2020b; Suá22: Suárez-Mascareño et al. 2022; Suá23: Suárez-Mascareño et al. 2023; Tol21: Toledo-Padrón et al. 2021; Tri18: Trifonov et al. 2018; Tri21: Trifonov et al. 2021; Zec19: Zechmeister et al. 2019.

First, CARMENES has succeeded in what it was designed for. Table 4 displays 82 planets discovered by CARMENES or with CARMENES contribution. The list is updated up to August 2024, and is sorted by publication date. Most of the planets, which were part of the results of the GTO, were already tabulated by Ribas et al. (2023). They enumerated 33 new planets and 26 confirmed planets from transiting candidate follow-up (Fig. 11). Ribas et al. (2023) also tabulated 17 planets that had been re-analysed with CARMENES data (e.g., Trifonov et al. 2018, 2020; Sarkis et al. 2018; Lalitha et al. 2019; Stock et al. 2020a; Amado et al. 2021; Cale et al. 2021; Radica et al. 2022). Table 4 does not show Barnard's Star "b", the planet candidate proposed by Ribas et al. (2018) that was afterwards challenged by Lubin et al. (2021) and Artigau et al. (2022). However, the list of others' exoplanets challenged by CARMENES is much longer, such as GJ 1151 "b" (proposed by Mahadevan et al. 2021 and ruled out by Perger et al. 2021 and Blanco-Pozo et al. 2023), TZ Ari "b" (proposed by Feng et al. 2020 and ruled out by Quirrenbach et al. 2022), or CN Leo "b" (proposed by Tuomi et al. 2019 and ruled out by Lafarga et al. 2021). See a full list of exoplanet candidates that turned to be a spurious signal, stellar rotation or alias, and their references, in Table 4 of Ribas et al. (2023). From all these results, the team has derived planet occurrence rates higher than one planet (1-1000 $M_\oplus$) per M dwarf (with orbital periods between 1 and 1000 d — Sabotta et al. 2021; Ribas et al. 2023). Their unique planet sample paved the way for statistical comparisons with planet formation theory, which revealed characteristic elements in the formation of planets around M dwarfs (Schlecker et al. 2022). Since most stars in the Milky Way are M dwarfs (Reylé et al. 2021), there are possibly more than 100 billion planets in our galaxy.

We describe below a few significant planet discoveries, either by CARMENES alone or in collaboration with other facilities on the ground and space.

- Teegarden's Star (Zechmeister et al. 2019; Dreizler et al. 2024). At merely 3.831 pc, Teegarden's Star is the 25th nearest star to the Sun (Reylé et al. 2021). It is an M7.0 V ultracool dwarf with a mass of only about 0.097 $M_\odot$. Together with TRAPPIST-1, it is the only planet host star at or below the 0.1 $M_\odot$ value. Planets b and c, with orbital periods and minimum masses of 4.90 d and 1.16 $M_\oplus$ and 11.42 d and 1.05 $M_\oplus$, respectively, were discovered first by Zechmeister et al. (2019) only with CARMENES data. With an instellation (insolation) of about 1.08 $S_\odot$ and a minimum mass only 16% larger than Earth's, Teegarden's Star b was considered the terrestrial planet with the highest Earth Similarity Index (Planetary Habitability Laboratory). Dreizler et al. (2024) used more CARMENES data, complemented with those from ESPRESSO, MAROON-X, and HPF for confirming and refining the parameters of planets b and c, and announcing a third planet, d, with an orbital period of 26.13 d and a minimum mass of only about 0.82 $M_\oplus$. Besides imposing stringent limits on the presence of transiting planets from TESS data and confirming the stellar rotation period at 96 d, Dreizler et al. (2024) also proposed a fourth terrestrial planet candidate for explaining a signal at 172 d. If confirmed, Teegarden's Star would resemble even more to the multi-planetary system TRAPPIST-1.
- GJ 357 (Luque et al. 2019). The transiting, hot, Earth-size planet GJ 357 b is optimal for atmospheric characterisation. It was one of the first great discoveries by NASA's TESS. Fortunately for the CARMENES team, who led the discovery work, GJ 357 was one of the GTO target stars and had enough data by the time of the TESS alert for displaying a powerful signal in the RV periodogram at 3.93 d, the period of the transiting planet b. After quickly adding more CARMENES and PFS (Planet Finder Spectrograph) data, together with the TESS light curves, archival RVs from HIRES, UVES and HARPS, and a careful host star investigation, Luque et al. (2019) were able to accurately determine the mass and radius of the planet, which at that time was the second closest transiting planet to the Sun. As a bonus, they also found two slightly more massive non-transiting planets, c and d, with orbital periods at 9.12 d at 78 d found only in the RV data. The outermost planet, GJ 357 d, is besides in its habitable zone.
- GJ 3512 (Morales et al. 2019). The host star is a 0.123-$M_{sol}$ M5.5 V star at less than 10 pc. The discovery of a giant exoplanet with a minimum mass of 0.463 $M_{Jup}$ orbiting such a very-low-mass star has since challenged planet formation models. Besides the large system's mass ratio, its long orbital period of 203 d and the presence of an additional companion at a much larger separation put constraints on the planet accretion and migration rates. Morales et al. (2019) concluded that disc instabilities may be more efficient in forming planets than previously thought. The discovery was based only on CARMENES data. A system analogous to GJ 3512 was discovered soon after by Quirrenbach et al. (2022), who reported the giant planet of $M \sin i$ = 0.213 $M_{Jup}$ around the M5.0 V-type star TZ Ari. To date GJ 3512 b and TZ Ari b are the only giant planets found around stars of less than 0.3 $M_\odot$ and, together with the GJ 912 B and GJ 3626 B brown dwarf candidates also discovered by CARMENES (Baroch et al. 2021), fill the gap between exoplanets around M dwarfs and M-dwarf double-line spectroscopic binaries.

- GJ 486 (Trifonov et al. 2021; Caballero et al. 2022). The GJ 486 (Wolf 437) system consists of a very nearby, relatively bright, weakly active M3.5 V star at just 8.08 pc with a warm transiting rocky planet of about 1.3 $R_\oplus$ and 3.0 $M_\oplus$. It is ideal for both transmission and emission spectroscopy and for testing interior models of telluric planets. It was discovered à la GJ 357 by Trifonov et al. (2021), who complemented CARMENES and TESS data with new, extremely precise MAROON-X RV data (Fig. 12). The planet GJ 486 b is not the "number one" in any list, but the combination of star's closeness (third closest transiting system), brightness (second brightest M dwarf with a transiting planet), extremely low activity, and visibility from both hemispheres, and planet's large transmission and emission spectroscopic metrics, short orbital period (of 1.47 d, which allows observing transits every three nights with a good time sampling), and a relatively high equilibrium temperature but below the limit for a molten surface make it unique. To all these advantages, Caballero et al. (2022) added a plethora of new observations of the star (optical interferometry with CHARA, X-ray and ultraviolet data with *XMM-Newton* and *Hubble*, ground-based high-resolution imaging and photometric monitoring) and simulations of the planet atmosphere and interior, which made GJ 486 b to be the best characterised Earth-mass planet outside the Solar System and one of the prime targets of *James Webb* (Fig. 13). Perhaps because of all of this, the International Astronomical Union included GJ 486 b in the NameExoWorld 2022 contest. Eventually, the star and planet received the approved names *Gar* and *Su*, which stand for Flame and Fire in Basque.
- GJ 1002 (Suárez-Mascareño et al. 2023) and Wolf 1069 (Kossakowski et al. 2023). Published almost simultaneously, these two stars completed the quintet of closest stars with Earth-mass planets in the habitable zone, together with Proxima Centauri, GJ 1061 and Teegarden's Star (also discovered by CARMENES). GJ 1002 hosts at least two planets, which were discovered from a fruitful collaboration between ESPRESSO and CARMENES, and Wolf 1069 hosts only one, which was discovered only with CARMENES data, but the three of them have minimum masses of 1.08-1.36 $M_\oplus$ and instellations sligthly smaller than unity, and orbit slowly-rotating M5.5 V dwarfs. In a sense, GJ 1002 b and c and Wolf 1069 can be considered a triplet (three offspring).
- GJ 12 (Kuzuhara et a. 2024). This system, announced simultaneously by two teams, was best characterised by a collaboration that included data obtained by IRD at Subaru, CARMENES and TESS. GJ 12 b is what many astronomers have been looking for decades: a nearby transiting temperate Earth-sized planet ideal for atmospheric transmission spectroscopy. Although the TRAPPIST-1 system also displays such main characteristics, the star GJ 12 is much brighter (4.2 mag in *G* and 2.7 mag in *J*), has an earlier spectral type and a much lower magnetic activity, and allows for a direct determination of the minimum mass from the Keplerian fit of its RV data (instead of from resonant chains and transit time variations). With an instellation of just 1.62 $S_\odot$ and an equilibrium temperature of ~315 K (assuming $A_{\rm Bond}$ = 0), and at a distance of 12 pc, GJ 12 b will soon became a prime target for *James Webb*.

**Table 5** Planets and atomic and chemical species identified with CARMENES [a].

| Planet | Species | Reference |
| --- | --- | --- |
| WASP-69 b | He I | Nortmann et al. 2018 |
|  | Na I, (Hα, K I, Ca II) | Khalafinejad et al. 2021 |
| HAT-P-11 b | He I | Allart et al. 2018 [b] |
| KELT-9 b | (He I) | Nortmann et al. 2018 |
|  | Hα | Yan & Henning 2018 |
|  | Ca II | Yan et al. 2019 |
|  | O I | Borsa et al. 2022 [b] |
|  | Ca II | Turner et al. 2020 [b] |
|  | Paβ | Sánchez-López et al. 2022b [b] |
|  | Si I, Mg I, Ca II, Fe I, (Al I, Ca I, Cr I…) | Ridden-Harper et al. 2023 [b] |
| GJ 436 b | (He I) | Nortmann et al. 2018 |
| HD 189733 b | He I | Salz et al. 2018 |
|  | $H_2O$ | Alonso-Floriano et al. 2019a |
|  | Na I, K I | Oshagh et al. 2020 |
|  | $H_2O$ | Cheverall et al. 2023 [b] |
|  | $H_2O$, ($CH_4$, CO, $H_2S$, HCN, $NH_3$) | Blain et al. 2024 [b] |
| HD 209458 b | He I | Alonso-Floriano et al. 2019b |
|  | $H_2O$ | Sánchez-López et al. 2019 |
|  | $H_2O$ | Sánchez-López et al. 2020 |
|  | (Hα, Na I, Ca II) | Casasayas-Barris et al. 2020 |
| HD 185603 b [c] | Hα, Na I Ca II, (Fe II) | Casasayas-Barris et al. 2019 [b] |
|  | Fe I, Fe II, Na I, Ca II | Nugroho et al. 2020 [b] |
|  | FeH? | Kesseli et al. 2020 [b,d] |
|  | Si I | Cont et al. 2022a |
|  | Fe I | Yan et al. 2022 |
|  | Hα, (He I) | Orell-Miquel et al. 2024 [e] |
| WASP-33 b | Ca II | Yan et al. 2019 |
|  | FeH? | Kesseli et al. 2020 [b,d] |
|  | Hα | Yan et al. 2021 |
|  | Fe I, TiO | Cont et al. 2021 |
|  | Si I | Cont et al. 2022a |
|  | Ti I, V I, OH, Ti II? | Cont et al. 2022b |
|  | (TiO, Ti I, V I) | Yang et al. 2024a |
|  | $H_2O$ | Yang et al. 2024b |
| GJ 3470 b | He I | Pallé et al. 2020 |
|  | ($H_2O$) | Dash et al. 2024 [b] |
| BD-02 5958 b,d [f] | (Hα, He I) | Carleo et al. 2021 [b] |
| WASP-76 b | Ca II, He I?, (Hα, Li I, Na I, K I) | Casasayas-Barris et al. 2021 |
|  | OH | Landman et al. 2021 [b] |
|  | $H_2O$, HCN | Sánchez-López et al. 2022a [b] |
| $\rho^{01}$ Cnc b | (HCN, $NH_3$, C2H2) | Deibert et al. 2021 [b] |
| HAT-P-32 b | Hα, He I | Czesla et al. 2022 |
| GJ 1214 b | He I? | Orell-Miquel et al. 2022 |
| HD 332231 b | (Hα, He I, Na I, K I, $H_2O$) | Sedaghati et al. 2022 |
| τ Boö Ab | (He I) | Zhang et al. 2020 [b] |
|  | $H_2O$, ($CH_4$, HCN, $NH_3$, C2H2) | Webb et al. 2022 [b] |
| HD 235088 b | He I | Orell-Miquel et al. 2023a |
|  | He I | Orell-Miquel et al. 2024 [e] |
| HAT-P-67 b | Ca II, Na I, Hα?, He I? | Bello-Arufe et al. 2023 [b] |
| WASP-12 b | Hα?, He I? | Czesla et al. 2024 |
| HD 201585 b [g] | Fe I, Ti I | Guo et al. 2024 |
| BD+65 902 d [g] | Hα, (He I) | Orell-Miquel et al. 2024 [e] |
| HD 149026 b | $H_2O$, (HCN) | Rafi et al. 2024 |

Notes. [a] Tentative detections are marked with a question mark, upper limits with parentheses. [b] Publications by teams different from the CARMENES consortium. [c] HD 185603 b has been named KELT-20 b and MASCARA-2 b by two competitor teams; here we use the Henry Draper identifier, as it is recommended by the International Astronomical Union. [d] Kesseli et al. (2020) also searched for FeH in the spectra of another ten CARMENES planetary systems. [e] Orell-Miquel et al. (2024) also searched for Hα and He I in the spectra of another nine planetary systems. [f] BD-02 5958 is sometimes named GJ 9827. However, the GJ numbers higher than 9000 were not assigned by Gliese or Gliese & Jahreiss, so here we use the much older Bonner Durchmusterung identifier. Carleo et al. (2021) investigated the two planets BD-02 5958 b and d. [g] HD 201585 and BD+65 902 have also been named MASCARA-1 and TOI-1136, respectively; here we use again the Henry Draper and Bonner Durchmusterung identifiers for homogeneity.

Second, CARMENES has also been able to advance in the observational study of atmospheres of transiting planets in both emission and transmission spectroscopy. Table 5 displays 20 systems with planets, mostly transiting, either gaseous or icy, with a number of atomic and chemical species identified (or tentatively detected) in their atmospheres by CARMENES. The Table displays individual atomic lines (hydrogen Hα and Paβ, helium He I $\lambda$1083 nm, sodium Na I $D_{1+2}$ $\lambda\lambda$589,590 nm, potassium K I $\lambda\lambda$767,770 nm, calcium Ca II $IRT_{a+b+c}$ $\lambda\lambda\lambda$850,854,866 nm), collection of many neutral metal lines (oxygen, silicon, iron, titanium, vanadium) and molecular bands (water, titanium oxide, hydroxil, iron hydride, hydrogen cyanide), with their respective references. Perhaps the most relevant result on this topic, which opened the window to more in-depth investigations of the atmospheres of exoplanets, was the detection in high spectroscopic resolution of He I in the infrared by Allart et al. (2018), Salz et al. (2018) and, especially, Nortmann et al. (2018). All this CARMENES observational effort has been supported with state-of-the-art theoretical work (Lampón et al. 2020, 2021a, 2021b, 2023).

**Table 6** CARMENES papers on exoplanet host star characterisation.

| Topic | Reference |
| --- | --- |
| Spectral type, activity, metallicity | Alonso-Floriano et al. 2015 |
| Close resolved multiplicity | Cortés-Contreras et al. 2017 |
| Spectroscopic binarity | Baroch et al. 2018 |
| Activity, line asymmetry | Fuhrmeister et al. 2018 |
| Activity, rotational velocity, spectroscopic binarity | Jeffers et al. 2018 |
| $T_{eff}$, log $g$, [Fe/H] | Passegger et al. 2018 |
| Activity, radial velocity | Tal-Or et al. 2018 |
| Photometric variability | Díez Alonso et al. 2019 |
| Activity, spectroscopic variability | Fuhrmeister et al. 2019a |
| Activity, spectroscopic variability | Fuhrmeister et al. 2019b |
| Activity, chromospheric model | Hintz et al. 2019 |
| $T_{eff}$, log $g$, [Fe/H] | Passegger et al. 2019 |
| Activity, spectral activity indicator | Schöfer et al. 2019 |
| Mass, radius | Schweitzer et al. 2019 |
| Magnetic field | Shulyak et al. 2019 |
| Photometric variability | Toledo-Padrón et al. 2019 |
| Spectroscopic abundance (Rb, Sr, Zr) | Abia et al. 2020 |
| Activity, convective shift | Baroch et al. 2020 |
| Luminosity, colour, spectral energy distribution | Cifuentes et al. 2020 |
| Activity, spectroscopic variability | Fuhrmeister et al. 2020 |
| Activity, chromospheric model | Hintz et al. 2020 |
| Activity, spectral activity indicator, radial velocity | Lafarga et al. 2020 |
| $T_{eff}$, log $g$, [Fe/H] | Passegger et al. 2020 |
| Spectroscopic binarity | Baroch et al. 2021 |
| Activity, spectroscopic variability | Johnson et al. 2021 |
| Activity, spectral activity indicator, radial velocity | Lafarga et al. 2021 |
| $T_{eff}$, log $g$, [Fe/H] | Marfil et al. 2021 |
| Activity, spectroscopic variability | Perdelwitz et al. 2021 |
| Spectroscopic abundance (V) | Shan et al. 2021 |
| Activity, spectroscopic variability | Fuhrmeister et al. 2022 |
| Activity, spectral activity indicator, radial velocity | Jeffers et al. 2022 |
| Activity, spectral activity indicator, radial velocity | Kossakowski et al. 2022 |
| [Fe/H] | Passegger et al. 2022 |
| Magnetic field | Reiners et al. 2022 |
| Activity, spectroscopic variability | Schöfer et al. 2022 |
| $T_{eff}$, [Fe/H] | Bello-García et al. 2023 |
| Activity, spectroscopic variability | Fuhrmeister et al. 2023a |
| Activity, spectroscopic variability | Fuhrmeister et al. 2023b |
| Activity, chromospheric model | Hintz et al. 2023 |
| Activity, radial velocity | Lafarga et al. 2023 |
| Activity, radial velocity | Perger et al. 2023 |
| Photometric variability | Shan et al. 2024 |
| $T_{eff}$, log $g$, [Fe/H] | Mas-Buitrago et al. 2024 |
| Spectroscopic abundance (Mg, Si) | Tabernero et al. 2024 |
| Kinematics, activity, age | Cortés-Contreras et al. 2024 |

And third, CARMENES has delved deeper into the topic whose motto is "*Know thy star, know thy planet*", i.e., the precise characterisation of the M dwarfs themselves. So, in order to prepare the CARMENES input catalogue and to carefully determine the astrophysical parameters of planet host stars, the CARMENES consortium has published a long series of papers on a number of stellar astrophysics topics, which are summarised in Table 6. They range from stellar activity (Hα and other prominent chromospheric lines, magnetic fields, flaring, activity indicators, rotational velocity, effect of activity on radial velocity measurements), through photometric variability, photospheric parameters ($T_{eff}$, log $g$, [Fe/H]), element abundances (Fe, Mg, Si, V, Sr, Zr, Rb), colours, absolute magnitudes and luminosities, to radii and masses.

But CARMENES, as a general-use instrument at the Calar Alto observatory, has also been used by other teams to study exoplanets or M dwarfs (Rajpurohit et al. 2018; Veyette & Muirhead 2018; Lillo-Box et al. 2022; Subjak et al. 2023 — see as well Table 4 for exoplanet atmosphere publications), as well as other exciting science. For example, CARMENES has been used to study stellar parameters of FGK-type stars (Marfil et al. 2020), variability of massive X-ray stars (Nazé et al. 2019), pulsations of early asymptotic-giant-branch stars (Začs & Pukītis 2023), radial-velocity photon noise as a function of spectral type (Reiners & Zechmeister 2020), and discovery of new broad interstellar absorption bands (Maíz Apellániz et al. 2021).

The CARMENES consortium is currently using the homonymous instrument for completing the radial-velocity survey for exoplanets around M dwarfs. This project, dubbed CARMENES Legacy+, will be running at the 3.5 m Calar Telescope until December 2025 together with other two legacy surveys, namely KOBE (K dwarfs observed with CARMENES) and CAVITY (galaxies in voids observed with PMAS). In parallel, radial-velocity follow up of TESS transiting planets, transmission and emission spectroscopy of exoplanet atmospheres and any other science are accomplished under open time observations with CARMENES, PMAS or the wide-field near-infrared imagers Omega-2000 and PANIC. Next, ESA's PLATO space mission will be launched at the end of 2026, and will likely look for transiting planets for at least three years in the southern hemisphere during its long duration pointing. By when ESA decides whether continuing in the south or looking for planets in the north during the "step-and-stare" or extension phases, for which CARMENES will be absolutely necessary, the super-integral field unit TARSIS should have arrived at the 3.5 m Calar Telescope and CARMENES will be almost 15 years old. Until then, and hopefully also afterwards, astronomers will make the most of one of the most successful spectrographs of the last two decades. It may actually happen that many years in the future the Habitable Worlds Observatory and the Large Interferometer For Exoplanets (or whatever name these space missions will eventually get) point to an Earth-like planet discovered by CARMENES and find the first incontrovertible biosignatures outside the Solar System. That would have been a good investment!

## Cross-references

- Extrasolar Planets: Change of Paradigm During the Twentieth Century
- The Naming of Extrasolar Planets
- ESPRESSO on VLT: An Instrument for Exoplanet Research
- HiCIAO and IRD: Two Exoplanet Instruments for the Subaru 8.2 m Telescope
- SPECULOOS Exoplanet Search and Its Prototype on TRAPPIST
- Exoplanet Research in the Era of the Extremely Large Telescope (ELT)
- Space Missions for Exoplanet Science: TESS
- Tools for Transit and Radial Velocity Modelling and Analysis
- Accurate Stellar Parameters for Radial Velocity Surveys
- Planets around Low-Mass Stars
- The "Spectral Zoo" of Exoplanet Atmospheres
- Factors Affecting Exoplanet Habitability

**Acknowledgements**
The authors wish to express their sincere thanks to all members of the CARMENES consortium for their hard and continuous work, and the Calar Alto staff for their expert support of the instrument and telescope operation. CARMENES is an instrument at the Centro Astronómico Hispano en Andalucía (CAHA) at Calar Alto (Almería, Spain), operated jointly by the Junta de Andalucía and the Instituto de Astrofísica de Andalucía (CSIC). CARMENES was funded by the Max-Planck-Gesellschaft (MPG), the Consejo Superior de Investigaciones Científicas (CSIC), the Ministerio de Economía y Competitividad and the European Regional Development Fund (ERDF) through projects FICTS-2011-02, ICTS-2017-07-CAHA-4, and CAHA16-CE-3978, and the members of the CARMENES consortium (Max-Planck-Institut für Astronomie, Instituto de Astrofísica de Andalucía, Landessternwarte Königstuhl, Institut de Ciències de l'Espai, Institut für Astrophysik Göttingen, Universidad Complutense de Madrid, Thüringer Landessternwarte Tautenburg, Instituto de Astrofísica de Canarias, Hamburger Sternwarte, Centro de Astrobiología and Centro Astronómico Hispano- Alemán), with additional contributions by the MINECO, the Deutsche Forschungsgemeinschaft through the Major Research Instrumentation Programme and Research Unit FOR2544 "Blue Planets around Red Stars", the Klaus Tschira Stiftung, the states of Baden-Württemberg and Niedersachsen, and by the Junta de Andalucía. We acknowledge financial support from the Agencia Estatal de Investigación (AEI/ 10.13039/501100011033) of the Ministerio de Ciencia e Innovación and the ERDF "A way of making Europe" through projects PID2022-137241NB-C4[1:4], PID2021-125627OB-C31, and the Centre of Excellence "Severo Ochoa" and "María de Maeztu" awards to the Instituto de Astrofísica de Canarias (CEX2019-000920-S), Instituto de Astrofísica de Andalucía (CEX2021-001131-S) and Institut de Ciències de l'Espai (CEX2020-001058-M). This article is dedicated to N. L. de Lacaille, J. Lalande, J. Kapteyn, E. E. Barnard, M. Wolf, F. E. Ross, G. van Biesbroeck, W. J. Luyten, O. Struve, P. van de Kamp, H. L. Giclas, W. Gliese, H. Jahreiss and all the astronomers that studied M dwarfs and dreamed of exoplanets before us.

# Figures

**Fig**. 1 Dome and building of the 3.5 m Zeiss telescope at Calar Alto on a snowy day during CARMENES assembly — Winter 2015-2016. Credit: CARMENES.

**Fig**. 2 The CARMENES logo. "solmirobauhaus" is the Spanish "Sol de Miró" as if it had been revisited by the German Bauhaus — Fall 2009. The logo is the 21st-century fusion of two very Spanish and German concepts. CARMENES was also the first astronomical instrument to have a soundtrack (listen to it in Antonio Arias' album *Multiverso II*). Credit: CARMENES.

**Fig**. 3 The CARMENES VIS channel optical layout — Spring 2012. The NIR one is very similar. Credit: CARMENES.

**Fig**. 4 The CARMENES wavelength coverage. *Top panel:* coadded, order-merged, channel-merged CARMENES VIS (blue) and NIR (red) template spectrum of exoplanet-host M3.5 V-type star GJ 486. Interruptions (grey areas) are due to strong telluric contamination and inter-order and NIR detector array gaps. *Bottom panels:* zoomed-in view of six representative, weakly magnetic-sensitive atomic lines. Black dashed lines are synthetic fits. Credit: CARMENES, adapted from Caballero et al. (2022).

**Fig**. 5 Installation of the CARMENES front-end at the Cassegrain focus of the 3.5 m Calar Alto telescope — Spring 2015. The light blue box and golden cylinders are the PMAS electronics and detector cryostats. Credit: CARMENES.

**Fig**. 6 Transportation of the wrapped CARMENES NIR channel tank inside the 3.5 m telescope building to its final location in the coudé room — Summer 2015. This was one of the scariest moments of the project, as the NIR tank contained during transportation the optical bench and some opto-mechanical mounts, unlike the VIS channel. The tough decision ended up being the right one. Credit: CARMENES.

**Fig**. 7 NIR optical bench after installing the camera, mounts of the collimator mirror and échelle grating, lower part of the radiation shield, and a number of temperature sensors (attached with kapton tape) — Summer 2015. The VIS optical bench is similar. Credit: CARMENES.

**Fig**. 8 NIR calibration unit optical bench after the Fabry-Pérot etalon — Spring 2024. Credit: CARMENES.

**Fig**. 9 The whole CARMENES data flow from the astronomer to the exoearth. Credit: CARMENES, adapted from Caballero et al. (2016b).

**Fig**. 10 Installation of the new joint VIS+NIR Fabry-Pérot etalon cryostat inside the 3.5m telescope calibration room — Spring 2024. From left to right: the NIR calibration unit

(partly rotated, red), silicone oil pump (cornered, partly visible), vaccum pump (cornered, red), new Fabry-Pérot cryostat (wrapped in aluminium and inside an isolated box, dubbed "Barbie") and optical bench with decommissioned VIS (blue) and NIR (red) Fabry-Pérot cryostats and halogen lamps (black). The VIS calibration unit and super-master-lamp storage tank are out of the image to the left and right, respectively. Credit: CARMENES.

**Fig**. 11 Scatter plots of the CARMENES first data release exoplanet sample compared to the complete sample of catalogued planets in the NASA Exoplanet Archive detected via RVs (903; small dots). Different symbols indicate planets newly detected from the CARMENES blind survey (33; stars), planets confirmed from transit follow-up (26; circles), and known planets re-analysed with CARMENES data (17; triangles). The three panels correspond to pairs of different relevant parameters, with the complementary colour scale introducing a third dimension. The histograms along the axes show distributions of the corresponding parameters for the CARMENES planet sample. The blue shaded band in the top-right panel represents the liquid-water habitable zone with limits defined by the 'runaway greenhouse' and 'maximum greenhouse' criteria. Credit: CARMENES, adapted from Ribas et al. (2023).

**Fig**. 12 Phase-folded transit and RV data and one-planet+Gaussian-process model fits. *Left:* light-curve model fit (black line) and CHEOPS+TESS data (CHEOPS transit #1: red, #2: orange, #3: yellow, #4: green, #5: light blue, #6: dark blue, #7: pink, and TESS: grey). *Right:* RV-curve model fit (black line with $\pm 1\sigma$ uncertainty marked with a grey shaded area) and CARMENES+MAROON-X data (CARMENES: green circles, MAROON-X Red: red symbols, and MAROON-X Blue: blue symbols. MAROON-X data are split into runs 1 [circles], 2 [squares], and 3 [triangles]). Error bars include original RV uncertainties (opaque) and jitter added in quadrature (semi-transparent). Credit: CARMENES, adapted from Caballero et al. (2022).

**Fig**. 13 Mass-radius diagram of all transiting exoplanets with mass determination (from RV or transit time variations) known in 2022, in comparison with the Solar System planets. Filled circles with error bars colour-coded by their host's $T_{eff}$ are planets with mass and radius uncertainties of less than 30%, and open grey circles are the others. The filled black star is Gl 486 b. Dashed coloured curves are theoretical models as specified in the legend. The Earth-like model is orange. The grey vertical dashed line is the deuterium burning mass limit at 13 $M_{Jup}$ ('planet'-brown-dwarf boundary). In the inset, we zoom in around the smallest planets and add mass-radius relationships informed by stellar abundances. We plot median and $1\sigma$ error regions following nominal relative abundances of Fe, Mg, and Si of the host star without (pink) and with (cyan) an empirical correction based on well-characterised super-Earths. The two outliers with very high densities and $M \sim 2.0\ M_\oplus$ are Kepler-1972b and c, which are two transiting planets with masses determined from transit time variations. Credit: CARMENES, adapted from Caballero et al. (2022).

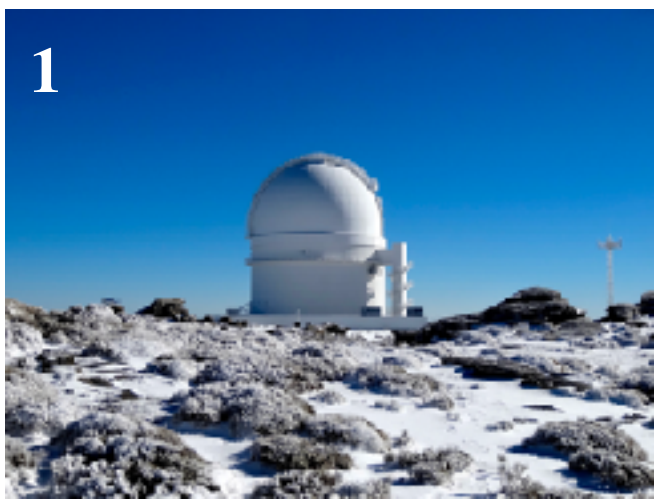 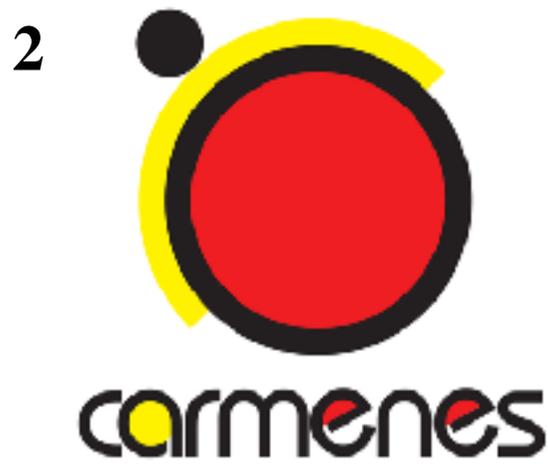 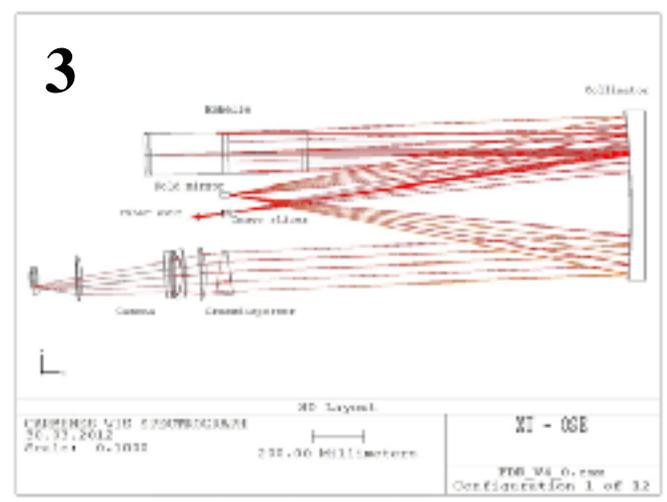 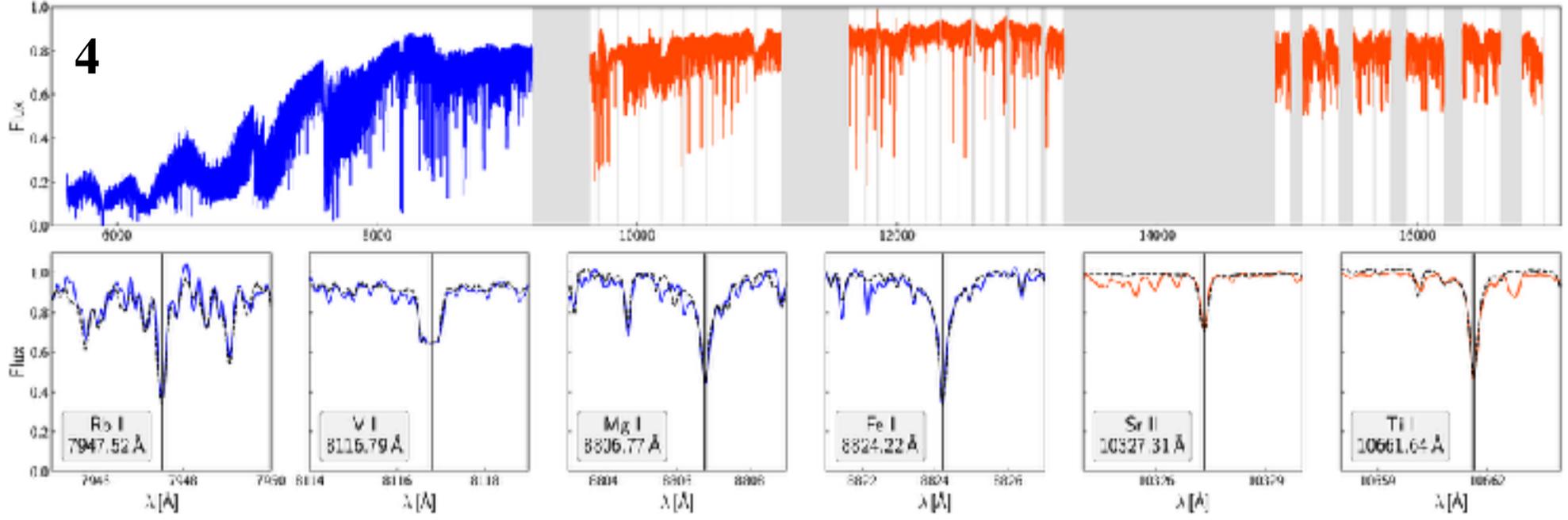 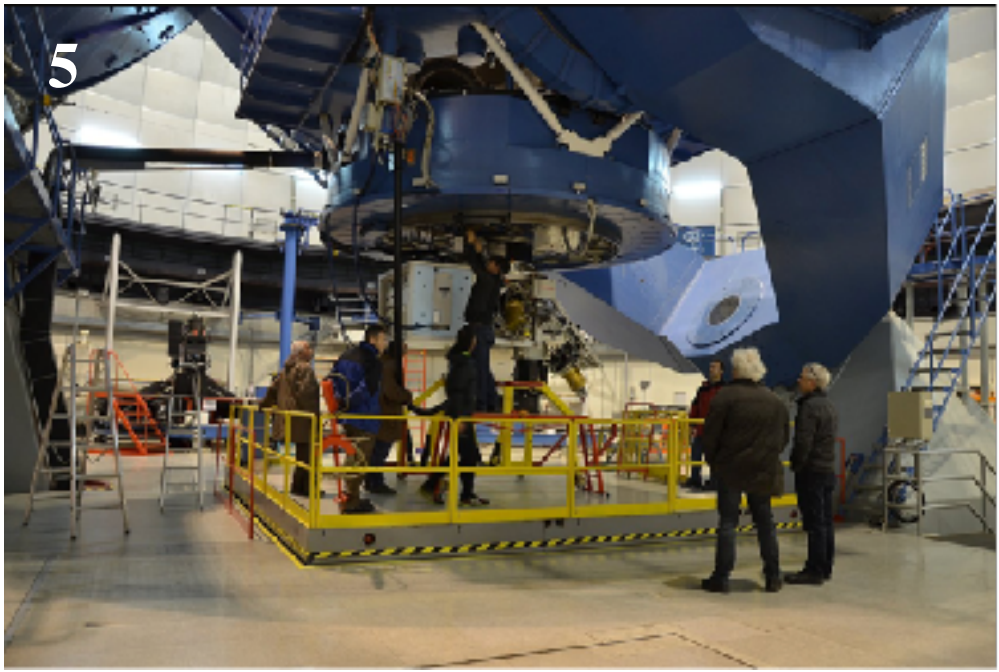 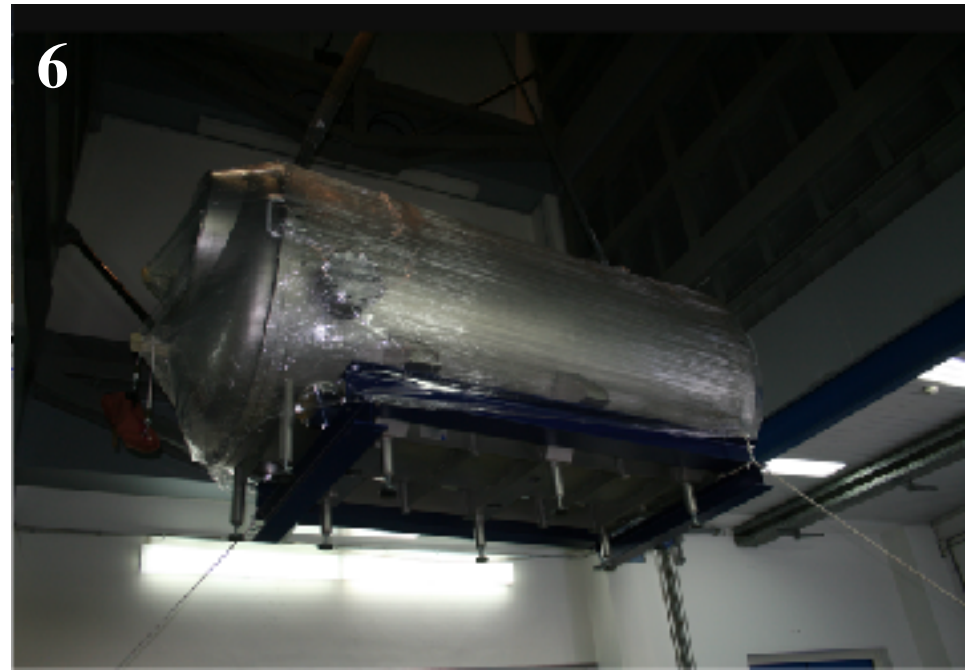 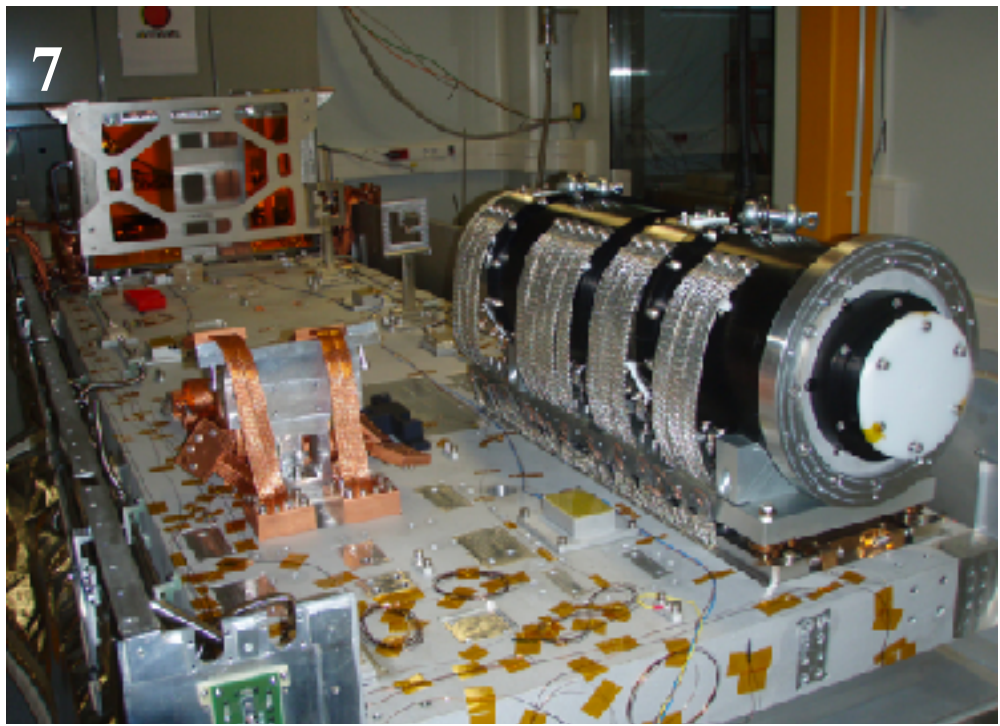 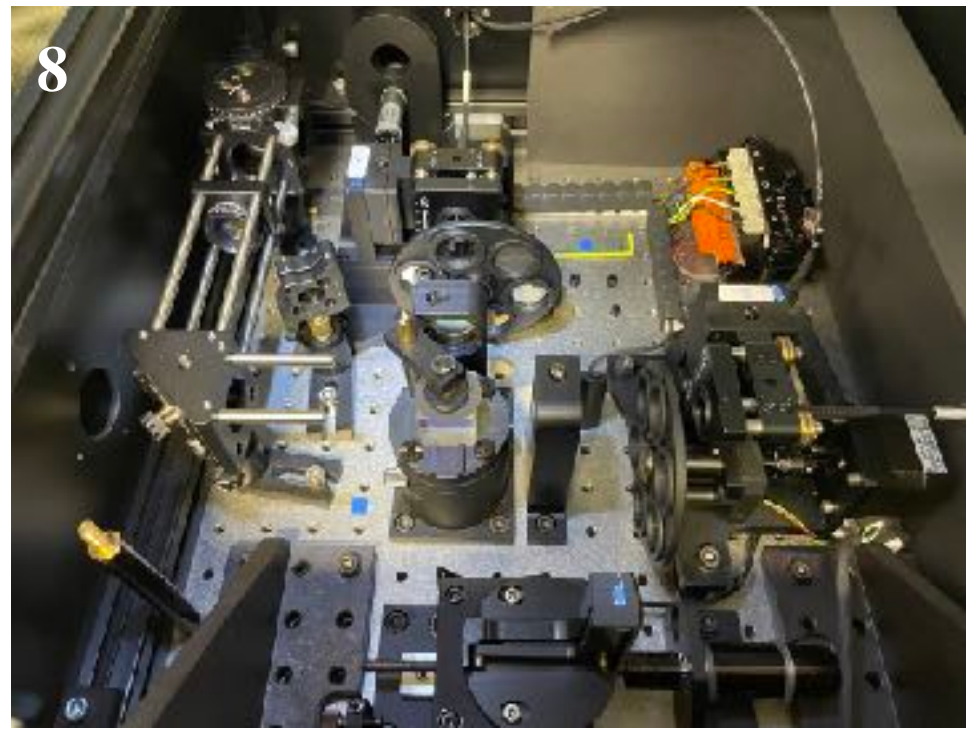

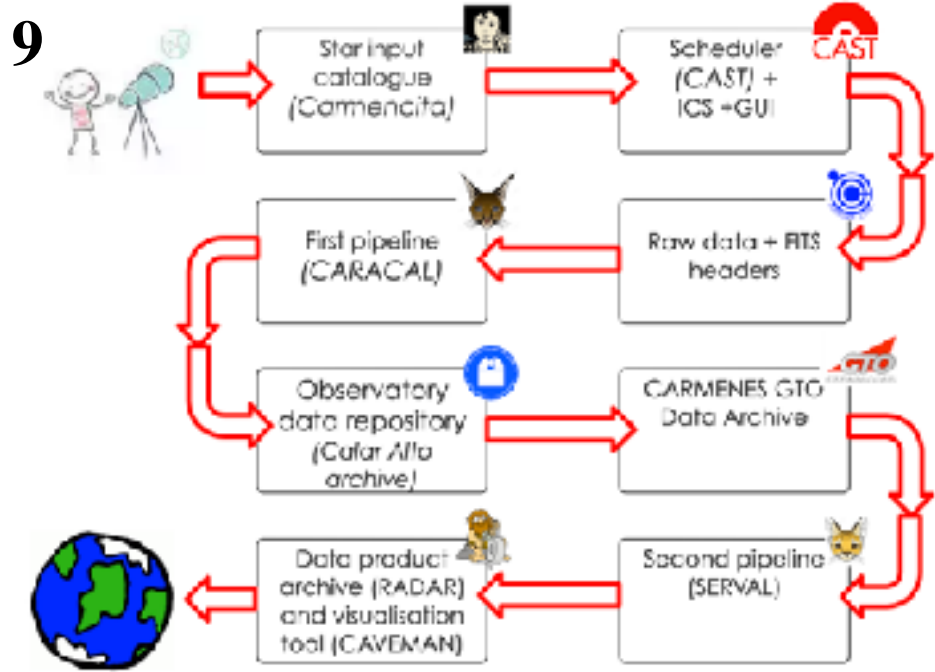

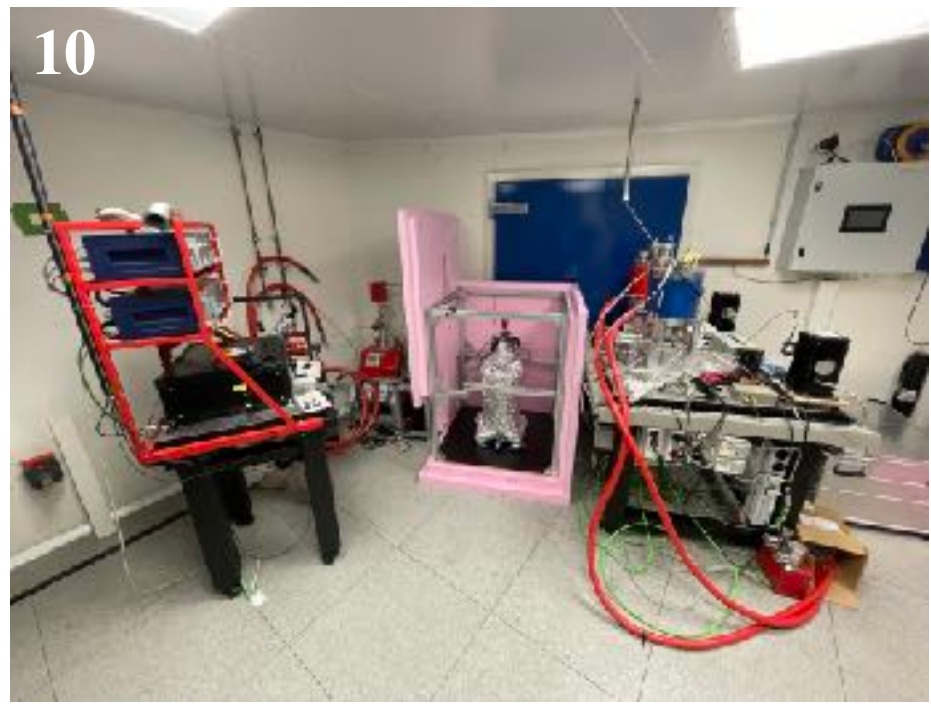

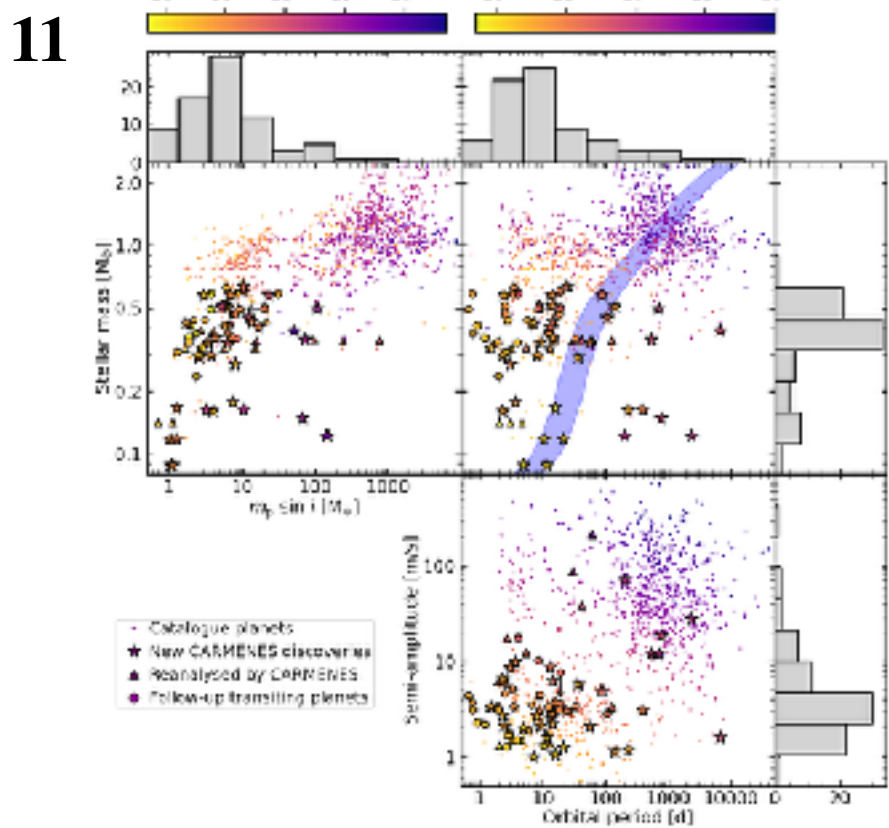

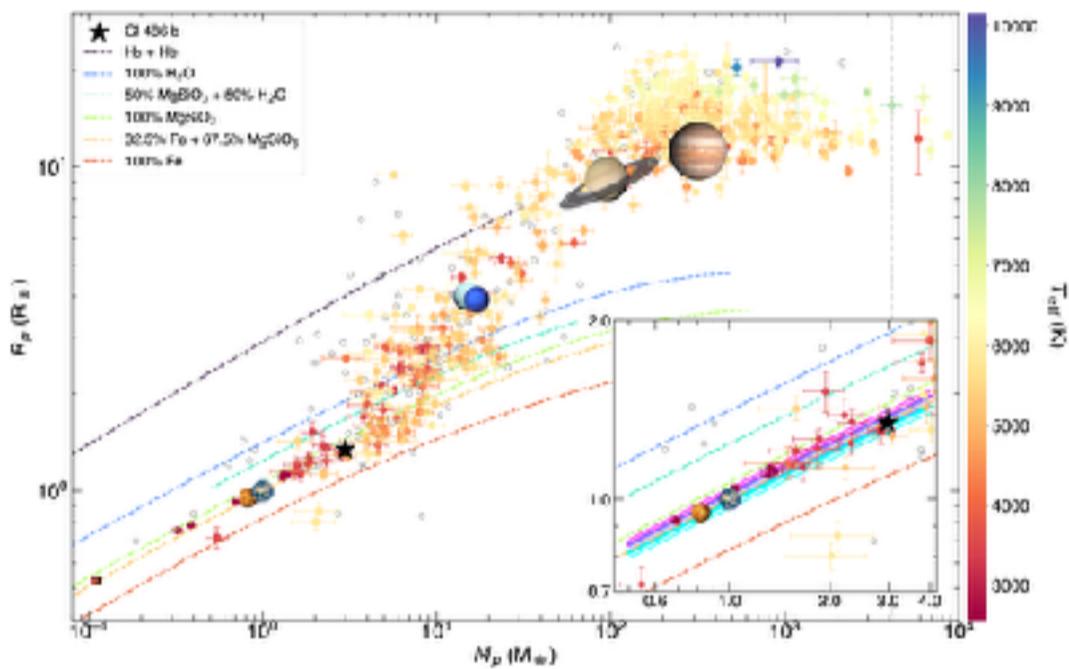

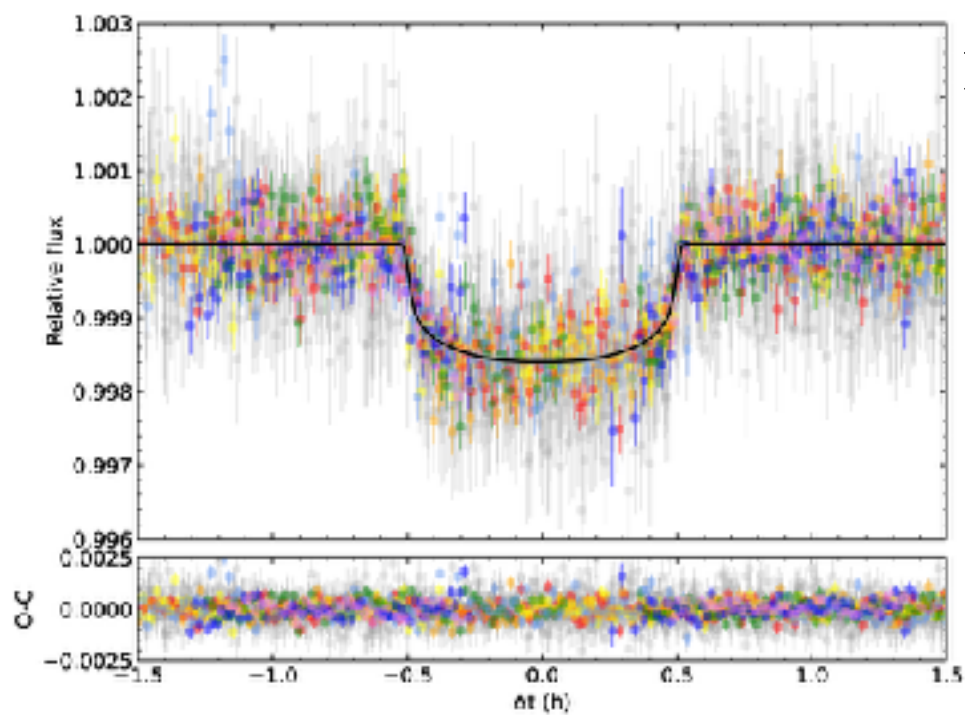

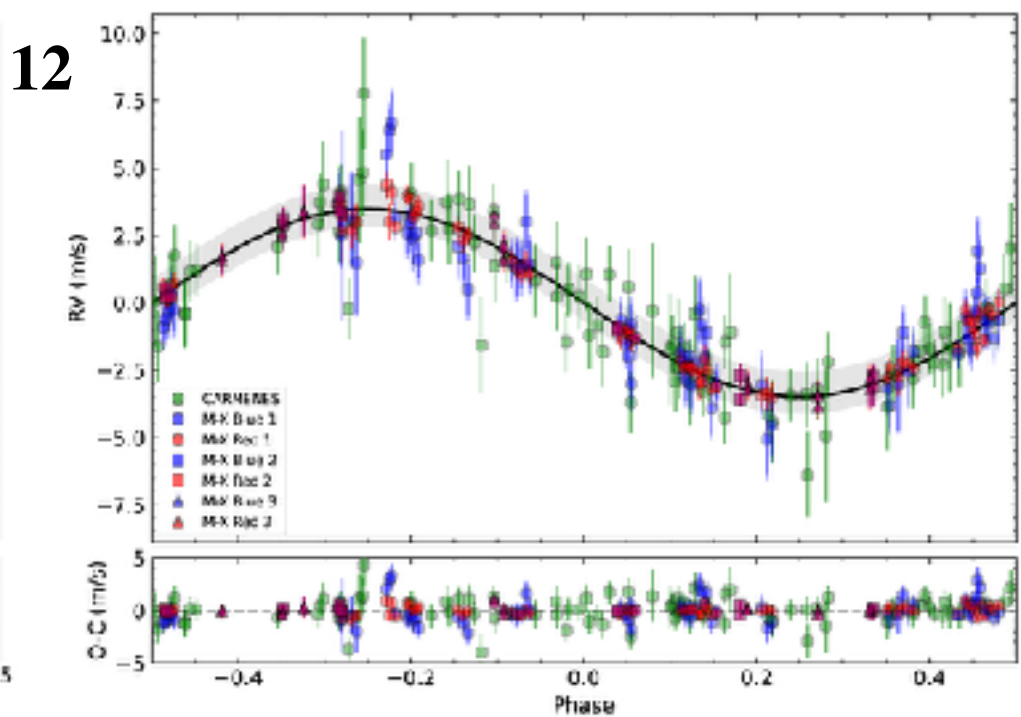

The corresponding images in full size can be downloaded at this link (massive file).